\documentclass[reprint, amsmath,amssymb, aps,prl]{revtex4-2}

\usepackage{graphicx}
\usepackage{dcolumn}
\usepackage{natbib}
\usepackage{bm}
\usepackage{braket}
\usepackage{hyperref}
\usepackage{amsthm}
\usepackage{xcolor}
\usepackage{cleveref}
\let\oldref\ref
\renewcommand{\ref}[1]{\pare{\oldref{#1}}}
\usepackage{amsfonts}

\begin{document}

\preprint{APS/123-QED}

\title{Topological superconductivity enhanced by exceptional points}

\author{R. Arouca}\email{rodrigo.arouca@physics.uu.se}
\author{Jorge Cayao}\email{jorge.cayao@physics.uu.se}
\author{Annica M. Black-Schaffer}\email{annica.black-schaffer@physics.uu.se}
\affiliation{Department of Physics and Astronomy, Uppsala University, Uppsala, Sweden}

\date{\today}

\begin{abstract}
Majorana zero modes (MZMs) emerge as edge states in topological superconductors and are promising for topological quantum computation, but their detection has so far been elusive. Here we show that non-Hermiticity can be used to obtain dramatically more robust MZMs. The enhanced properties appear as a result of an extreme instability of exceptional points to superconductivity, such that even a vanishingly small superconducting order parameter already opens a large energy gap, produces well-localized MZMs, and leads to strong superconducting pair correlations. Our work thus illustrates the large potential of enhancing electronic ordering, here in the form of topological superconductivity, using non-Hermitian exceptional points.  
\end{abstract}
\maketitle

Topological superconductors are currently intensively pursued since they can host Majorana zero modes (MZMs), which are promising for topological quantum computation due to their non-Abelian statistics \cite{beenakker2020search, sarma2015majorana}. The simplest model to present topological superconductivity is the Kitaev chain \cite{ kitaev2001unpaired, beenakker2013search,aguado2017majorana, flensberg2021engineered, tanaka2011symmetry, PhysRevB.56.892,PhysRevB.53.R11957,PhysRevLett.110.117002, pikulin2012topological, wang2015spontaneous, yuce2016majorana, zeng2016non,klett2017relation, li2018topological, kawabata2018parity, mcdonald2018phase,avila2019non, lieu2019non,cayao2020odd, li2020coalescing, maiellaro2020non, zhou2020non, wang2021unconventional, jiang2021non, liu2021fate, cao2021universal, rahul2021topological, zhao2021defective, PhysRevB.103.134507, PhysRevB.103.224207, PhysRevB.87.235421}. While being a simple model composed of spinless fermions with $p$-wave pairing, it is still effectively realized at interfaces between conventional superconductors and topological insulators \cite{PhysRevB.79.161408,PhysRevLett.103.107002,sacepe2011gate,PhysRevB.87.220506,banerjee2018signatures,culcer2020transport}, semiconductors \cite{lutchyn2010majorana, oreg2010helical,mourik2012signatures, PhysRevLett.108.147003,PhysRevB.96.205425,cayao2018andreev,cayao2018finite}, and even using  magnetic impurity chains \cite{PhysRevB.84.195442, nadj2013proposal, pientka2013topological,nadj2014observation, PhysRevB.88.180503,PhysRevB.94.100501,PhysRevB.102.104501, PhysRevLett.111.147202,PhysRevLett.111.186805,pawlak2016probing}. 

In the ideal situation, MZMs emerge at zero energy and are well separated from the quasicontinuum of other states by a topological gap, which creates an isolated subspace for the topological qubit. However, in practice, the topological gap strongly depends on material properties and is usually smaller than the superconducting order parameter in the parent superconductor. This considerably limits a realistic Majorana-based computation \cite{sarma2015majorana, sarma22}, which urgently calls for platforms with much larger values of the topological gap.

With the advent of non-Hermitian quantum mechanics, it has been predicted that system properties can be drastically enhanced due to the strong sensitivity of non-Hermitian systems \cite{wiersig2014enhancing,hodaei2017enhanced, chen2017exceptional, miller2017exceptional,wiersig2020review}. This sensitivity is due to the presence of non-Hermitian degeneracies, also known as exceptional points (EPs), which are points where both eigenvalues and eigenstates coalesce \cite{berry2004EP, moiseyev2011non, heiss2012EP, kato2013perturbation, ghatak2019new,xu2016topological, san2016majorana,hodaei2017enhanced,wang2019arbitrary,miri2019exceptional, yoshida2019symmetry,okuma2019topological,okugawa2019exceptional,budich2019symmetry,yoshida2019ER,yoshida2019ER_2,kawabata2019classification,yang2019fermion,li2020topological,denner2020exceptional, bergholtz2019exceptional, mandal2021symmetry, stalhammar2021classification, aquino2020exceptional, lourencco2021non, para2021probing, cayao2022exceptional, schafer2022symmetry, ashida2020non, delplace2021symmetry}. The sensitivity gets even larger when increasing the number $N$ of coalescing energy levels, where $N$ is also referred to as the order of the EP. In fact, for a dimensionless perturbation $x$, the spectrum changes as $x^{1/N}$ \cite{wiersig2014enhancing,hodaei2017enhanced, chen2017exceptional, miller2017exceptional,wiersig2020review}, revealing that even an infinitesimally small $x$ can induce a sizable effect for sufficiently large $N$. A tantalizing question arises here if it may be possible to exploit this strong sensitivity of non-Hermitian systems to enhance the topological gap protecting the MZMs. 

In this work, we show that non-Hermiticity can dramatically enhance topological superconductivity, producing much more robust MZMs. In particular, we consider the non-reciprocal Kitaev chain and show that the presence of a high-order EP in the normal state makes the system highly unstable to superconductivity, such that even a vanishingly small superconducting order parameter is sufficient to open a sizable gap protecting the MZMs, resulting in exceptionally enhanced topological superconductivity.
With sizable superconductivity being hard to achieve experimentally, our results open a compelling avenue for designing systems with much more robust MZMs.

\begin{figure}[!tb]
    \centering
    \includegraphics[width=0.75\linewidth]{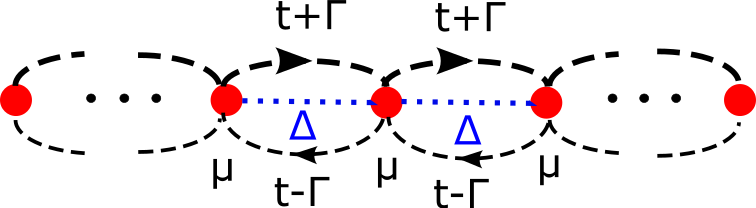}
    \caption{ Lattice representation of the HNK Hamiltonian with lattice sites (red dots) and terms connecting different sites (dashed/dotted lines).}
    \label{fig_lattice}
\end{figure}

\textit{Model and its normal state}.---\label{sec_HNK} To investigate and most clearly elucidate exceptionally enhanced topological superconductivity, we consider a non-Hermitian extension of the simplest known topological superconductor: the Kitaev chain. Further, as we want to investigate systems with strong non-Hermitian sensitivity, we seek a very simple model where the normal state presents a high-order EP. 
An ideal candidate for the latter is the Hatano-Nelson (HN) model \cite{hatano1996localization, gong2018topological, bergholtz2019exceptional}, where simple non-reciprocal hoppings give rise to high order EP \cite{hatano1996localization, gong2018topological, bergholtz2019exceptional}. Combining the two models we arrive at the Hatano-Nelson-Kitaev (HNK) chain, which is illustrated in Fig.~\ref{fig_lattice} and given by
    \begin{align}
        H = & -\sum\limits_{r}\left[\left(t+\Gamma\right)c^\dagger_{r+1}c_{r}+\left(t-\Gamma\right)c^\dagger_{r}c_{r+1}+\mu c^\dagger_{r}c_{r}\right] \nonumber \\
        & + \sum\limits_{r}\left( \Delta c^\dagger_{r+1}c^\dagger_{r}+{\rm H.c.}\right),
        \label{eq_HNK}
     \end{align}
where $c^\dagger_r$ ($c_r$) creates (annihilates) a spinless fermion at site $r$, $t$ is the nearest neighbor hopping, $\mu$ the chemical potential, $\Delta$ the $p$-wave superconducting order parameter, and $\Gamma$ encodes the non-reciprocity.

The Hermitian, or Kitaev, limit, found at $\Gamma=0$, hosts a topological phase for $\left|\mu/t\right|<2$, characterized by a bulk topological invariant and MZMs at the endpoints of any finite chain \cite{kitaev2001unpaired, beenakker2013search,aguado2017majorana, flensberg2021engineered}. 
Inclusion of the non-reciprocal $\Gamma$ into the Kitaev chain has already been considered \cite{cao2021universal, rahul2021topological, zhao2021defective}. Even though non-Hermitian systems can present new topological phases or extend the topological phases in Hermitian systems \cite{kawabata2019symmetry, bergholtz2019exceptional}, the topological phase diagram for the HNK chain was found to remain, being characterized by the non-Hermitian version of the same invariant \cite{cao2021universal} and showing localized MZMs. It has additionally been shown that the non-Abelian braiding is preserved in non-Hermitian systems \cite{san2016majorana, ezawa2018higher}. These earlier results seem to indicate that one does not gain much from making the Kitaev chain non-Hermitian. However, an entirely overlooked aspect is how the normal-state EPs can dramatically enhance the topological superconducting phase, albeit the phase diagram boundaries do not change.

Before showing that the EP in the normal state spectrum enhances the superconducting phase, we need to describe this EP. The normal state of the HNK chain in just the HN model \cite{hatano1996localization, gong2018topological, bergholtz2019exceptional} or  Eq.~\eqref{eq_HNK} with $\Delta=0$. Due to the non-reciprocal hopping, the HN model presents remarkably different properties depending on the boundary conditions. In the presence of periodic boundary conditions (PBC), it hosts non-Hermitian topological phases but no EPs. In contrast, the open boundary conditions (OBC) system does not have a topological phase but has an EP of the order of the system size $L$ \cite{hatano1996localization, gong2018topological, bergholtz2019exceptional}. As we are interested in the EP effects and MZMs, we focus on OBC in the main text but discuss the PBC system in the Supplementary Material (SM) \cite{SM} (see also Refs.~\cite{sanchez2012transfer,fu2022degeneracy,li2022exact,feinberg1997non,sukhachov2020non, datta1997electronic} therein). In general, for non-Hermitian systems, the bulk spectrum for OBC can be captured by the non-Bloch Hamiltonian $h$, which takes values in a complex generalization of the Brillouin zone \cite{yao2018edge, lee2019anatomy, lee2019unraveling}. In many cases, including the HN model, we can then obtain an analytical expression for the eigenvalues of $h$ and, consequently, of the OBC spectrum \cite{yao2018edge, lee2019anatomy, lee2019unraveling}. However, for the HNK chain, to the best of our knowledge, there exists no analytical solution, see SM \cite{SM}. Therefore, we are here forced to primarily rely on a numerical solution of the Boguliubov-de Gennes (BdG) Hamiltonian \cite{de2018superconductivity, zhu2016bogoliubov}, where $H=\mathbb{C}^\dagger \mathbb{H}_{\rm BdG}\mathbb{C}$, using the Nambu vector $\mathbb{C}=\big(\begin{matrix}c_0&\cdots&c_{L-1}&c_0^\dagger&\cdots c_{L-1}^\dagger \end{matrix}\big)^T$, for an open chain with $L$ sites. 
    \begin{figure}[!tb]
        \centering
        \includegraphics[width=\linewidth]{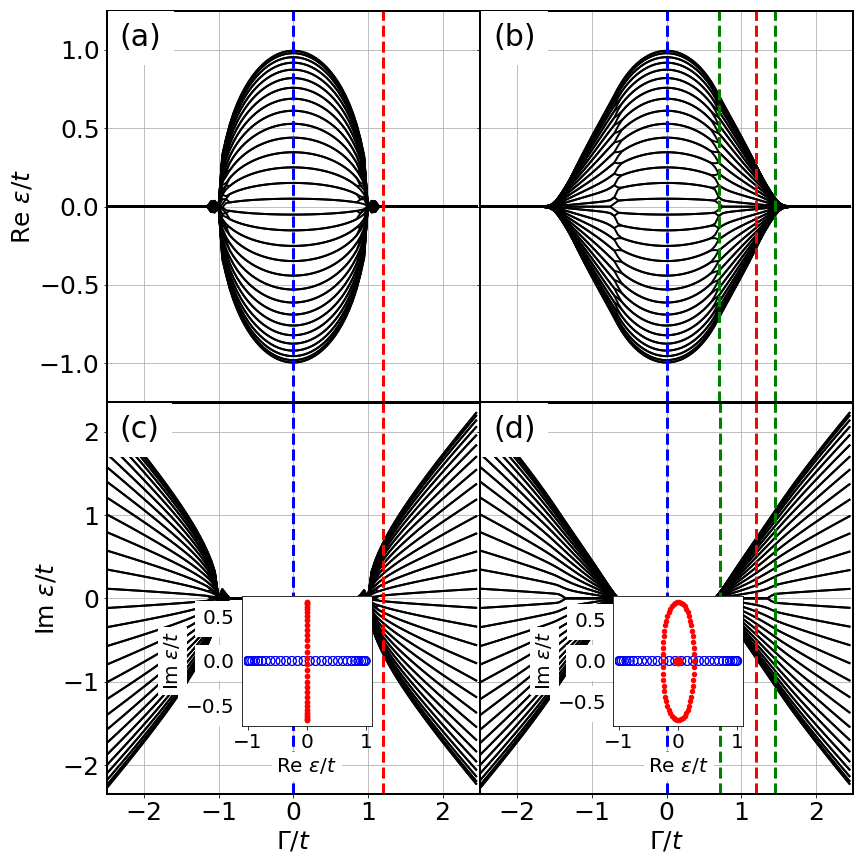}
        \caption{Spectrum of the OBC HNK system as a function of $\Gamma/t$ with $\Delta=0$ (a,c) and $\Delta=10^{-10} t$ (b,d). First (second) row shows the real (imaginary) part of the energy spectrum. Insets in (c,d) show the energy levels in the complex plane in the Hermitian limit ($\Gamma=0$, blue dashed lines and open circles) and in the $\mathcal{PH}$-broken phase ($\Gamma=1.2 t$, red dashed lines and filled circles). Green dashed lines mark the region with MZMs. Other parameters: $\mu=0$, $L=30$. }
        \label{fig_spectrum_Delta}
    \end{figure}
The real/imaginary parts of the spectrum of Eq.~\eqref{eq_HNK} in the normal state are shown in Fig.~\ref{fig_spectrum_Delta}(a)/(c) as a function of $\Gamma/t$. For $\left|\Gamma/t\right|<1$, the energies are completely real, while for $\left|\Gamma/t\right|>1$, the energies are purely imaginary. The presence of regions with completely real energies in the system is due to pseudo-Hermitian ($\mathcal{PH}$) symmetry, while the developing of a spectrum with complex conjugated pairs marks the spontaneous breaking of this symmetry \cite{kunst2019non, alvarez2018non, longhi2020non, longhi2020_nonbloch, gardas2016non}. These $\mathcal{PH}$-preserved and $\mathcal{PH}$-broken phases are separated by an EP of the order of the system size, where \textit{all} energy levels coalesce \cite{hatano1996localization, gong2018topological, bergholtz2019exceptional}
\footnote{In the non-Bloch formalism, this complete coalescence can be understood as a non-Bloch band collapse \cite{kunst2019non, alvarez2018non, longhi2020non, longhi2020_nonbloch, lee2019unraveling}, where the imaginary part of the momentum diverges for all modes, such that the energies accumulate at a single value, see e.g.~Ref.~\cite{arouca2021non} for a discussion in the HN model}. 
\begin{figure*}[!tb]
        \centering
        \includegraphics[width=\linewidth]{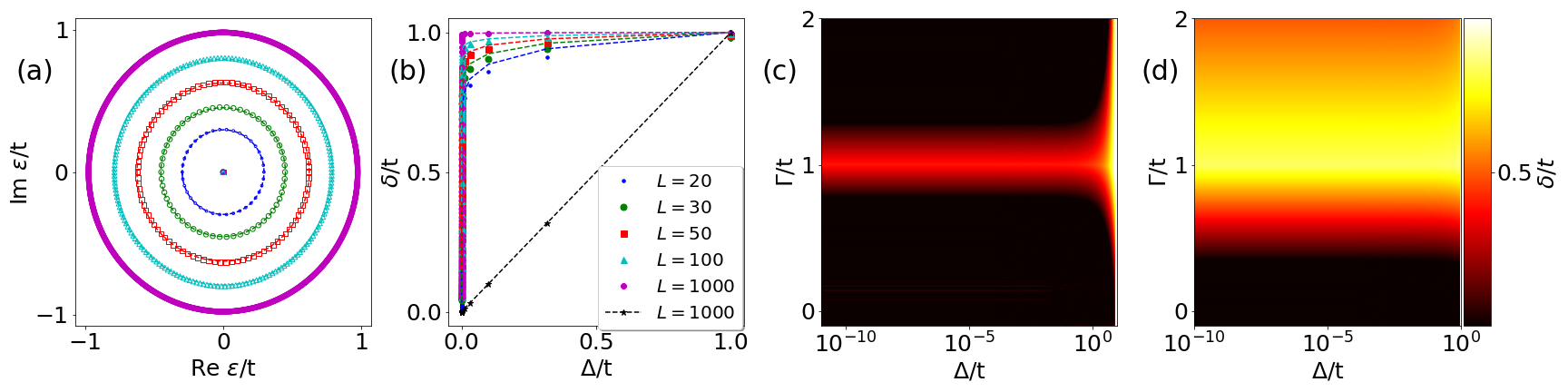}
        \caption{Exceptional sensitivity of the spectrum $\epsilon$ (a) and minigap $\delta$ (b) for $\Gamma=t$ or $\Gamma=0$ at multiple different systems sizes $L$, using $\Delta=10^{-10} t$. Points represent numerical data, the dashed lines analytical results (Eqs.~\eqref{eq_epsilon_an}). Minigap $\delta$ as a function of $\Delta/t$ and $\Gamma/t$ for $L=30$ (c) and $L=100$ (d). Other parameters: $\mu = 0$.}
        \label{fig_epsilon}
    \end{figure*}
\textit{Exceptional gap}.---Having understood the HNK system without superconductivity and its EP, we next add a small superconducting order parameter $\Delta$. Doing so, we arrive at a surprising conclusion: already adding a vanishingly small $\Delta$ dramatically changes the energy levels. In Fig.~\ref{fig_spectrum_Delta}(b)/(d), we show the real/imaginary part of the spectrum for the extremely small $\Delta=10^{-10} t$ (see SM  \cite{SM} for complementary data at different $L$). For such small values of $\Delta$, most of the spectrum barely changes from the non-superconducting case. The exception is for $\Gamma$-values around the normal-state EP, which disappears and is replaced by a continuum of bands, which joins pairwise in multiple EPs. In addition, there are also two zero energy states, two MZMs, clearly isolated from the rest of the other modes, easily seen when plotting the complex energies (red filled circles) in the inset of Fig.~\ref{fig_spectrum_Delta}(d). We further notice that although these MZMs are clearly isolated modes, there still exist bulk modes with, separately, Re$(\epsilon)=0$ or Im$(\epsilon)=0$. Thus, one may naively think the system is gapless, just as in the Hermitian limit (blue circles), but thanks to the complex energy spectrum, there exists a finite region in the complex plane around the MZMs where no bulk states are found. This spectrum configuration corresponds to a point gap, which provides a non-Hermitian topological protection of the MZMs \cite{bergholtz2019exceptional}.  As a consequence, even for a vanishingly small $\Delta$, we find a finite minigap $\delta=\left|\epsilon_{\rm QP}-\epsilon_{\rm{MZM}}\right|$, with $\epsilon_{\rm QP}$ being the lowest quasi-particle mode, protecting the MZMs. The finite minigap exists for a finite range of $\Gamma$, marked by the vertical green dashed lines in Fig.~\ref{fig_spectrum_Delta}(b,d).
Although we find no general analytical expression for the minigap protecting the MZM, we can gain understanding by perturbatively adding superconducting pairing at the normal-state EP. The details are given in the SM \cite{SM} and here we focus on the results. At $\Gamma/t=1$, the (bulk) energy levels $\epsilon_m$ are given, to the lowest order in $\Delta/t$, by 
\begin{equation}
   \epsilon_m/t\approx \, \exp\left({i  \pi m\frac{L}{2L-2}}\right) \sqrt[L-1]{\Delta/t},
    \label{eq_epsilon_an}
\end{equation} 
with $m=0, 1,\dots, 2L-2$.
From this result, we see that for sufficiently large $L$, $\sqrt[L-1]{\Delta/t}\sim 1$ for any $\Delta$, which directly explains the exceptional sensitivity of the energy levels to any perturbation $\Delta$. This result is in agreement with the recently reported sensitivity of non-Hermitian systems \cite{kato2013perturbation, wiersig2022response}, but they were not considering superconductivity or other electronic ordering. This approximate energy level expression also clarifies why the energies are complex since each of them is a point on a circle in the complex plane. The radius of this circle also automatically gives the minigap, $\delta/t\approx \left(\Delta/t\right)^{1/(L-1)}$ since the MZMs are at zero energy.
In Fig.~\ref{fig_epsilon}(a,b), we compare these perturbative and thus approximate analytical results for the energy levels (a) and minigap (b) with the earlier obtained exact numerical results using $\Gamma/t=1$ and $\Delta/t=10^{-10}$ for many different values of $L$. We find that analytical (lines) and numerical (points) results agree so well with each other that they are not distinguishable in the figure. We also directly see that as $L$ increases, the radii of the energy level circles increase, leaving a larger minigap. In particular, we notice that the $(L-1)$-root expression in Eq.~\eqref{eq_epsilon_an} means that there is a power-law behavior of the minigap $\delta$, as we show in the SM \cite{SM}. This is very different from the Hermitian limit (black points, lines) where the minigap is always $\delta\sim \Delta$. We thus establish, both numerically and analytically, that the normal-state EP makes the system exceptionally unstable in the thermodynamic limit to superconductivity, generating a finite minigap even for vanishingly small $\Delta$. It is the coalescence of all levels at the normal-state EP that generates this extreme instability as a collective phenomenon. 

The above analytical analysis is only valid at the normal-state EP at $\Gamma/t=1$, but the instability towards superconductivity extends for a finite range of $\Gamma$ around the EP. We illustrate this in Figs.~\ref{fig_epsilon}(c,d), where we plot the minigap $\delta$ as a function of both $\Gamma$ and $\Delta$ for both $L=30$ (c) and $L=100$ (d). The strong system size dependence is also visible here, as a sizable minigap is present in a much larger region for the larger $L$ system, clearly emphasizing the collective aspect of the effect. The spectrum is also affected by other values of the parameters, but MZMs are still present in a large region of the parameter space. For instance, although we here only report data for $\mu=0$, for $\Delta=10^{-10} t$ the enhancement appears in the large regime $|\mu/t| \lesssim 1$. To not deviate from the main result, we discuss these details in the SM \cite{SM} and in Supplementary Videos (SVs) \cite{SV}.

Taken together, we find that already a vanishingly small $\Delta$ around the normal-state EP changes the features of the spectrum in the complex plane, opening a sizable point gap that produces isolated MZMs. We coin this remarkable sensitivity {\it exceptionally enhanced topological superconductivity}. The sensitivity is a collective effect, arising due to the accumulation of all system modes at the EP and thus growing with system size. This can be viewed as a non-Hermitian equivalent of how the accumulation of density of states in flat band systems creates superconductivity, as well as other electronic ordered states, as recently illustrated in e.g.~twisted bilayer graphene at the magic angle \cite{Cao2018, Cao2018SC, Balents2020, Andrei2020} or bi- and trilayer graphene in electric fields \cite{Zhou2021trilayer, Zhou2022bilayer}.

\begin{figure}[!tb]
        \centering
        \includegraphics[width=1.1\linewidth]{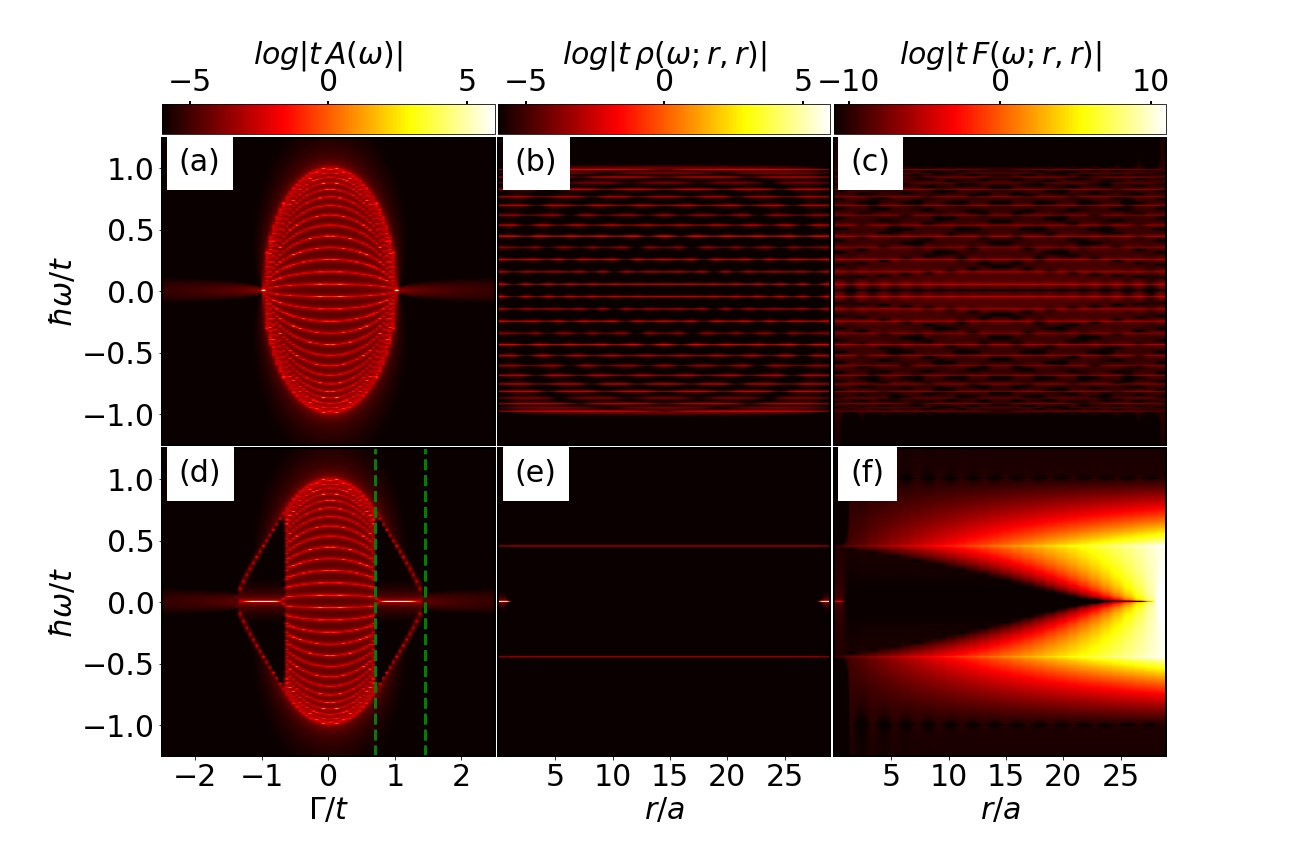}
        \caption{Spectral weight $A$ (a,d), LDOS $\rho$ (b,e), and local superconducting pair correlations $F$ (c,f) as a function of $\Gamma$, $\omega$ and position $r$ along the chain ($a$ is the lattice parameter). First column shows $A$ for $\Delta=0$ (a) and $\Delta=10^{-10}t$ (d), second column shows $\rho$ for $\Delta=10^{-10} t$ and $\Gamma=0$ (b) and $\Gamma=t$ (e) , and the third column shows $F$ for $\Delta=10^{-10} t$ and $\Gamma=0$ (c) and $\Gamma=t$ (f). Other parameters: $L=30$, $\mu=0$.}
        \label{fig_E_Delta}
    \end{figure}

\textit{Robust MZMs and superconductivity}.---
Exceptional topological superconductivity is not restricted to influencing the topological gap, but next we show how it also manifest in the spectral weight and superconducting pair correlations.  These are encoded in the Nambu Green's function $\mathcal{G}$ \cite{marino2017quantum, kozii2017non, sukhachov2020non, cayao2022exceptional} \footnote{We include a small ($10^{-7}t$) imaginary positive term in $\hbar \omega$ in the definition of $\mathcal{G}$ to obtain finite linewidths for modes with zero imaginary part.} 
    \begin{equation}
         \mathcal{G}(\omega)=\frac{1}{\hbar \omega \hat{1}-\mathbb{H}_{\rm BdG}}=\left(\begin{matrix}G_e&F_{eh}\\F_{he}&G_h\end{matrix}\right),
        \label{eq_G}
    \end{equation}
where $G_e$ and $G_h$ are the normal electron and hole Green's functions, respectively, while $F_{eh}$ and $F_{he}$ are the anomalous contributions containing the superconducting pair correlations or amplitudes.
From $\mathcal{G}$, we obtain all two-particle observables of the system, see SM \cite{SM} for details. First, we calculate the spectral weight $A(\omega)=-\text{Im Tr} \left[\mathcal{G}(\omega)-\mathcal{G}^\dagger(\omega)\right]/(2\pi)$,
which represents the density of states at the frequency (energy) $\omega$ ($\hbar\omega$). Note that Tr here denotes the sum over both electron and hole components and spatial coordinates. In Figs.~\ref{fig_E_Delta}(a,d), we plot $A$ for the HNK system. At $\Delta=0$ (a), $A$ is only finite when $\mathcal{PH}$-symmetry is preserved, and we find mainly non-zero real energy values. The only visible peak occurs at the normal-state EP ($\left|\Gamma/t\right|=1$) due to its accumulation of modes. For a vanishingly small $\Delta=10^{-10} t$ (d), the same behavior, with $A$ reflecting the non-zero real part of the spectrum, is now only found in a narrower region of $\Gamma$. Instead, there is a huge peak at $\omega=0$ that extends until the real energies disappear. This is the same region, marked by green dashed lines in Fig.~\ref{fig_spectrum_Delta}, where we find a point gap separating the MZMs from the remaining modes. Thus, this peak in $A$ is nothing else than a zero bias peak (ZBP) produced by the MZMs \cite{aguado2017majorana, flensberg2021engineered}. 
Moreover, since the sizable point gap is associated with finite imaginary energies, which broadens the poles for all modes but the MZMs, we find a very clear ZBP with no other low-energy spectral weight, again signaling the exceptional enhancement of the topological phase.

The exceptional enhancement is not only seen in the spectrum and spectral weight but also in the spatial properties of the system. To illustrate this, we consider the local density of states (LDOS), $\rho(\omega;r)=-\text{Im}\left[G_{e}(\omega; r, r)\right]/\pi$.
In Fig.~\ref{fig_E_Delta}(b), we plot $\rho$ as a function of position along the chain for $\Delta=10^{-10} t$ in the Hermitian limit ($\Gamma=0$) and find, as expected, essentially uniform occupation across energies and position. In contrast, tuning to the normal-state EP at $\Gamma=t$ in Fig.~\ref{fig_E_Delta}(c), we find very pronounced and well-localized peaks at zero energy at each end of the chain. This demonstrates that non-Hermiticity also strongly promotes MZM localization, which, together with the large minigap, establishes the existence of robust MZMs.
Farther away from the EP, we still find MZMs, but they then decay further into the bulk, see SV \cite{SV}, since the coherence length of the MZMs is proportional to $\delta^{-1}$ \cite{aguado2017majorana, flensberg2021engineered}. 

Finally, we also show that the superconducting pair correlations, expressed through $F_{eh}$, are also exceptionally enhanced due to the normal-state EP. Keeping a vanishingly small $\Delta=10^{-10} t$, we find in the Hermitian limit in Fig.~\ref{fig_E_Delta}(c) that the local pair correlations are of the same order as $\Delta$ and delocalized in the lattice, as expected. In contrast, for $\Gamma=t$ in Fig.~\ref{fig_E_Delta}(f), we find an excessively strong $F$, up to 20 orders of magnitude larger than $\Delta$. Such a strong response shows that, although the superconducting order parameter $\Delta$ is vanishingly small, there exists strong superconducting pairing, further cementing the notion of exceptionally enhanced topological superconductivity. Another remarkable feature is the localization of $F_{eh}$ at the right edge due to the Non-Hermitian skin effect \cite{yao2018edge,lee2019anatomy, lee2019unraveling,borgnia2020non, PhysRevLett.124.086801, zhang2020correspondence, li2020critica, helbig2020generalized, xiao2020non, kunst2018biorthogonal,yao2018edge,yokomizo2019non,lee2018tidal,song2019non,okuma2019topological, PhysRevLett.124.086801,yang2019auxiliary, yoshida2020mirror}. If instead plotting $F_{he}$, we find a similar localization at the left edge, illustrating that electron and hole components localize on opposite edges.
    
\textit{Concluding remarks}.---We show that even a vanishingly small superconducting order parameter $\Delta$ in a non-reciprocal Kitaev chain can create exceptional topological superconductivity with robust MZMs, due to a normal-state EP of the order of the system size. The robustness of the MZMs stems from the opening of an exceptionally enhanced point gap in the system, which both generates the minigap protecting the MZMs and strongly localized MZMs. Moreover, the superconducting pair correlations are also exceptionally enhanced. While the most dramatic behavior occurs for vanishingly small $\Delta$ close to the normal-state EP, any non-Hermiticity still enhances MZM robustness and superconductivity. Our results are also stable to changes in the chemical potential. We expect that this relative insensitivity to parameter values also makes the results stable to weak disorder. 
In terms of a direct experimental realization, cold atom systems are promising as they can host both $p$-wave pairing \cite{diehl2011topology, buhler2014majorana} and non-reciprocity \cite{gong2018topological, li2020topological}. Our results are also transferable to other superconductors as they only require the opening of a point gap close to an EP in the normal state. Additionally, we speculate that the same reasoning can be applied to other correlated phases that are amplified by state degeneracies, simple examples being the Stoner mechanism for ferromagnetism \cite{stoner1938collective} and the Peierls instability \cite{pouget2016peierls}. Finally, an interesting outlook is to understand the role of particle-hole symmetry in exceptionally enhanced superconductivity, since the enhancement of the gap with a power $1/(L-1)$, while preserving MZMs, seems to be related to the analysis of perturbations of EPs in the presence of chiral and sublattice symmetry in Ref.~\cite{sayyad2022realizing}.

\begin{acknowledgments}
We are grateful for discussions with E.~J.~Bergholtz, J.~C.~Budich, and T.~Yoshida.
We acknowledge financial support from the Knut and Alice Wallenberg Foundation and the European Research Council (ERC) under the European Union Horizon 2020 research and innovation programme (ERC-2017-StG-757553). J.~C.~acknowledges financial support from the Swedish Research Council (Vetenskapsr\aa det Grant No.~2021-04121).
\end{acknowledgments}

\bibliographystyle{apsrev4-2}
\bibliography{non_herm}

\begin{onecolumngrid}
	
%==============Supplemental material=============
\begin{center}
	{\fontsize{12}{12}\selectfont
		\textbf{Supplementary Material: Topological superconductivity enhanced by exceptional points\\[5mm]}}
	{\normalsize {R. Arouca, Jorge Cayao, and Annica M. Black-Schaffer\\}
	{\small \it Department of Physics and Astronomy, Uppsala University, Uppsala, Sweden}\\[0.5mm]}
\end{center}
\section{Periodic boundary conditions and lack of exceptional points}

In the main text, we show that for open boundary conditions (OBC), the presence of a high order exceptional point (EP) in the normal state Hatano-Nelson-Kitaev (HNK) chain (i.e.~just the Hatano-Nelson (HN) model) leads to an extreme, or exceptional, sensitivity to superconductivity. Here we show that due to the absence of such an EP for periodic boundary conditions (PBC), there exists no such enhancement of superconductivity in the periodic version of the HNK chain. With PBC, the Hamiltonian is block-diagonal in momentum space and reads
\begin{equation}
    H=\mathbb{C}^\dagger \mathbb{H}_{\rm BdG}\mathbb{C}=\frac{1}{2}\sum\limits_{k} \left(\begin{matrix}c^\dagger_k& c_{-k}\end{matrix}\right)h_{\rm BdG}(k)\left(\begin{matrix}c_k\\ c^\dagger_{-k}\end{matrix}\right),
\end{equation}
such that we can write the BdG Hamiltonian as
\begin{equation}
h_{\rm BdG}(k)=\left(\begin{matrix}-\mu-2t \cos(ka)-2i\Gamma\sin(ka)&-2i\sin(ka) \Delta\\ 2i\sin(ka) \Delta&\mu+2t \cos(ka)-2i\Gamma\sin(ka)\end{matrix}\right).
\end{equation}
From here, we obtain the energy spectrum directly 
\begin{equation}
    \epsilon_\pm(k)=\pm \sqrt{\left[\mu+2t \cos(ka)\right]^2+4\Delta^2\sin(ka)^2}-2i\Gamma \sin(ka).
\end{equation}
The real part of the spectrum is equal to the energy spectrum of the (Hermitian) Kitaev chain, and the imaginary part is just linear in $\Gamma$.
The Green's function, in this case, is also easy to obtain and is given by
\begin{eqnarray}
    \mathcal{G}(\omega, k)&=&\frac{1}{\hbar \omega\hat{1}-h_{\rm BdG}(k)}=\frac{1}{B(\omega, k)}\left(\begin{matrix}\hbar\omega+2i \Gamma \sin(ka)-\left[\mu+2t\cos(ka)\right]&-2i \Delta \sin(ka)\\2i \Delta \sin(ka)&\hbar\omega+2i \Gamma \sin(ka)+\mu+2t\cos(ka) \end{matrix}\right)
\end{eqnarray}
where $B(\omega, k)=\hbar^2\omega^2+2i \hbar\omega\Gamma \sin(ka)-4(\Delta^2+\Gamma^2)\sin^2(ka)-\left[\mu+2t\cos(ka)\right]^2$. We note that in this case, for $\Delta$ small, $F_{eh/he}\propto \Delta$, directly illustrating that the superconducting amplification seen for OBC in the main text is always absent for PBC. 

The fact that the periodic system is barely modified by a small $\Delta$ is also clearly seen when we consider how the spectrum and the spectral weight change with $\Delta$, as illustrated in Fig.~\ref{fig_spectrum_Delta_PBC}. This figure is to be compared with Fig.~2 of the main text, which shows the same quantities with similar parameters, but with OBC.  For $\Delta=0$, we seem to have a continuum of bands in both the real, Fig.~\ref{fig_spectrum_Delta_PBC}(a), and imaginary, Fig.~\ref{fig_spectrum_Delta_PBC}(d), part of the spectrum. However, the analysis of the spectrum in the complex plane, inset of Fig.~\ref{fig_spectrum_Delta_PBC}(d), shows that the system has a point gap, but now without any Majorana zero modes (MZMs) in the middle, as expected for a periodic system. This point gap does not reflect any interesting properties in the spectral weight,  see Fig.~\ref{fig_spectrum_Delta_PBC}(e), aside from the van Hove singularities at $\hbar \omega=\pm 2t$, related to the borders of the Brillouin zone, and the spectrum of the PBC Kitaev chain at $\Gamma=0$.  The inclusion of a vanishingly small $\Delta=10^{-10}t$ does not alter the spectrum, see Figs.~\ref{fig_spectrum_Delta_PBC} (b,e) or the spectral weight, see Fig.~\ref{fig_spectrum_Delta_PBC}(h). For a sufficiently large $\Delta=0.1 t$, we find an opening of a small gap in the real part of the energy in Fig.~\ref{fig_spectrum_Delta_PBC}(c), but with an unchanged imaginary energy part, see Fig.~\ref{fig_spectrum_Delta_PBC}(f), which leads to line gap of a similar size as $\Delta$ for all $\Gamma$, see inset of Fig.~\ref{fig_spectrum_Delta_PBC}(f).

\begin{figure}[!htb]
        \centering
        \includegraphics[width=\linewidth]{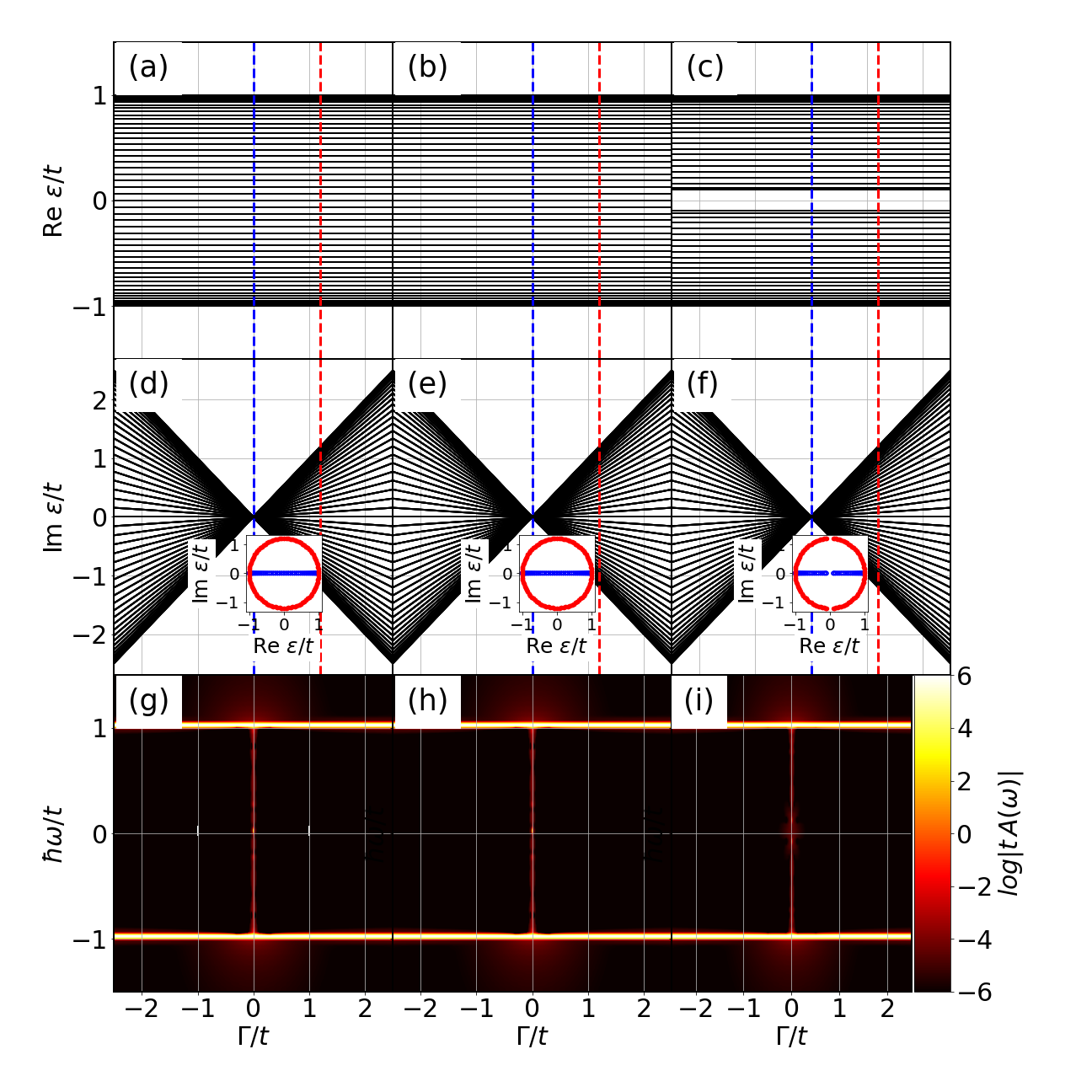}
        \caption{
        Spectrum and spectral weight of the PBC HNK system as a function of $\Gamma/t$ (same as Fig.~2 in the main text but for PBC). (a,d,g) have $\Delta=0$, $(b,e,f) \Delta=10^{-10} t$, and (c,f,j) $\Delta=0.1 t$, where the first (second) row shows the real (imaginary) part of the spectrum, while the last row shows the log of the spectral weight $\left|A \right|$. Insets of (d,e,f) show the energy levels in the complex plane for the system in the Hermitian limit ($\Gamma=0$, blue dashed lines and open circles) and in the $\mathcal{PH}$-broken phase ($\Gamma=1.2 t$, red dashed lines and filled circles). Other parameters: $100$ $k$-points (corresponding to a lattice with $L=100$ sites) and $\mu=0$.}
        \label{fig_spectrum_Delta_PBC}
    \end{figure}

To summarize, the abrupt change of the electronic properties, even for a vanishingly small value of the superconducting order parameter $\Delta$ observed in the main text (see Figs.~2, 3, and 4 of the main text) for OBC is clearly not present for PBC. 
This is manifest both in the superconducting pair correlations, obtained analytically, and the numerical analysis of the spectrum and the spectral weight. The fact that the HNK system with PBC is not exceptionally sensitive to superconducting pairing can be traced back to the lack of EPs in the spectrum for the normal state (HN model).

\clearpage
\section{Non-Bloch formalism}\label{sec_lit}

In this section of the SM we discuss the non-Bloch formalism and motivate why it is difficult to obtain analytical results for the HNK chain. Some general properties of this model, and of non-Hermitian superconductors in general, are explained in detail in Ref.~\cite{cao2021universal}. Here, we reproduce some of these results for pedagogical reasons. In non-Hermitian systems, there is a breaking of the traditional bulk-boundary correspondence, such that it cannot be used. Nevertheless, there exist different approaches to obtaining the spectrum and the states, such as the transfer matrix \cite{sanchez2012transfer, kunst2019non}, and the non-Bloch formalism \cite{yao2018edge, lee2019anatomy, lee2019unraveling}. In particular, the non-Bloch formalism makes that the $\exp(i k )$ of the Fourier transform is substituted by a complex number $z$. Using it, the non-Bloch Hamiltonian of the HNK chain is of the form \cite{cao2021universal}
\begin{equation}
    h(z)=\Gamma\left(z-z^{-1}\right)\hat{1}+i \Delta\left(z-z^{-1}\right)\sigma_y+\left[\mu+t\left(z+z^{-1}\right)\right]\sigma_z.
\end{equation}
The secular equation $\det\left[\epsilon\hat{1}-h(z)\right]=0$ then becomes a polynomial equation in $\epsilon$ and $z$. In our case
\begin{equation}
    \det\left[\epsilon\hat{1}-h(z)\right]=\left[\epsilon-\Gamma\left(z-z^{-1}\right)\right]^2-\left[\mu+t\left(z+z^{-1}\right)\right]^2+\Delta^2\left(z-z^{-1}\right)^2=0,
\end{equation}
Therefore, the energies are given by
\begin{equation}
    \epsilon_\pm(z)=\Gamma\left(z-z^{-1}\right)\pm\sqrt{\left[\mu+t\left(z+z^{-1}\right)\right]^2-\Delta^2\left(z-z^{-1}\right)^2},
\end{equation}
while the secular equation is a fourth-order polynomial equation for $z$
\begin{equation}
    \left[\epsilon-\Gamma\left(z-z^{-1}\right)\right]^2-\left[\mu+t\left(z+z^{-1}\right)\right]^2+\Delta^2\left(z-z^{-1}\right)^2=0=\left[\epsilon z-\Gamma\left(z^2-1\right)\right]^2-\left[\mu z+t\left(z^2+1\right)\right]^2+\Delta^2\left(z^2-1\right)^2.
\end{equation}
The four solutions for $z$ can be sorted according to their absolute value. The physical solutions are given by the solutions $z_2$ and $z_3$, which satisfy $|z_2|=|z_3|$. In some cases, like the HN model, $|z|$ does not depend on $\epsilon$. In this case, we can write $z=\exp{\left[i k(\epsilon)\right]}|z|$ and obtain an effective Bloch Hamiltonian called a surrogate Hamiltonian. However, for the HNK chain, $|z|$ is a function of $\epsilon$, $|z|(\epsilon)$,  such that the OBC modes can only be obtained numerically. Even though there exist no analytical solutions for the bands, $z(\epsilon)=z^{-1}(-\epsilon)$ due to particle-hole symmetry \cite{cao2021universal}, leading to electron and holes presenting a non-Hermitian skin effect (NHSE) in different edges of the system. 
The fact that $|z|(\epsilon)$ explains why we in the main text have to resort primarily to a numerical approach in our description of exceptionally enhanced superconductivity, although we can still derive key properties of the energy levels and minigap analytically. 

\clearpage

\section{Finite size effects}
In this section of the SM we extend the analysis presented in Figs.~2(b,d) of the main text to show how the spectrum at vanishingly small $\Delta = 10^{-10}t$ changes for different sizes. In particular, Fig.~\ref{fig_spectrum_Delta_L} shows the spectrum and spectral weight for the OBC HNK system with $\Delta=10^{-10} t$ for increasing values of $L$ from left to right using $L =10,20,30,40,50$. We clearly see how the exceptional sensitivity of the non-superconducting HN system to the introduction of superconductivity heavily depends on system size as the normal-state EP becomes more and more degenerate with an increasing number of lattice sites. As such it is a collective effect due to the accumulation of all the system modes at the EP in the normal state. 
In Supplementary Video (SV) 1, we continuously vary $L$ to even more clearly illustrate this effect.

\begin{figure}[!htb]
        \centering
        \includegraphics[width=\linewidth]{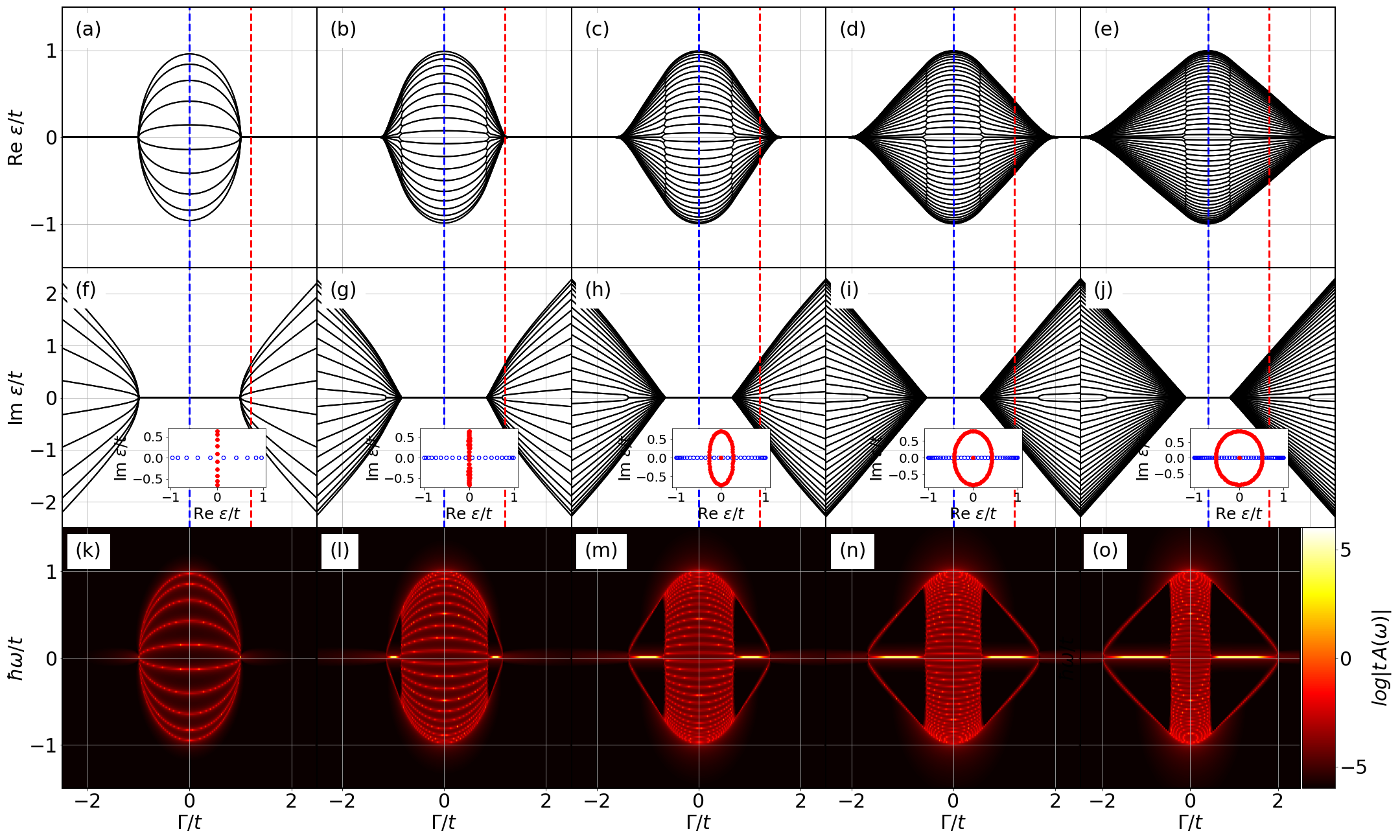}
        \caption{Spectrum and spectral weight of the OBC HNK system as a function of $\Gamma/t$ for $L=10$ (first column, (a,f,k)), $L=20$ (second column, (b,g,l)), $L=30$  (third column, (c,h,m)), $L=40$  (fourth column, (d,i,n)), and $L=50$  (fifth column, (e,j,o)). The first (second) row shows the real (imaginary) part of the spectrum, while the third row shows the log of the spectral weight $|A|$. 
      Insets in (f-j) show the energy levels in the complex plane for the system in the Hermitian limit ($\Gamma=0$, blue dashed lines and open circles) and in the $\mathcal{PH}$-broken phase ($\Gamma=1.2 t$, red dashed lines and filled circles). Other parameters: $\Delta=10^{-10} t$ and $\mu=0$. }
        \label{fig_spectrum_Delta_L}
    \end{figure}

The exceptional sensitivity of the non-superconducting HN system to the introduction of superconductivity is a collective effect due to the accumulation of all the system modes at an EP in the normal state.  Therefore, it is a phenomenon that depends heavily on system size as the EP becomes more and more degenerate with an increasing number of lattice sites.

\subsection{Lattice with an odd number of sites}
In both the main text and in the rest of this SM, we consider lattices with an even number of lattice sites because systems with an odd number of lattice sites can have spurious zero modes that are \textit{not} MZMs. This is due to the combination of $\mathcal{PH}$-symmetry, which constraints the energies to appear in $(\epsilon, \epsilon^*)$ pairs, and particle-hole symmetry, which causes the energies to appear in $(\epsilon, -\epsilon)$ pairs, which make lattices with an odd number of sites to always host zero energy modes. These modes can be trivial, if degenerate with spatially overlapping eigenstates, or isolated MZMs. In Fig.~\ref{fig_spectrum_Delta_L_odd}, we show a version of Fig.~\ref{fig_spectrum_Delta_L}, i.e.~we display the spectrum and spectral weight of the system for increasing system sizes, but for lattices with an odd number of sites, still keeping $\Delta=10^{-10} t$. Overall, Figs.~\ref{fig_spectrum_Delta_L} and \ref{fig_spectrum_Delta_L_odd} are quite similar. But for the odd-numbered lattices, we find an additional  ZBP for all $\Gamma/t$, which clearly illustrates the spurious, non-MZM, zero-energy modes in odd-numbered systems. We also note that there are no new EPs points induced by superconductivity in contrast to the even-numbered systems. However, the normal-state EP of the HN model is still present, which is the root of the exceptionally enhanced superconductivity, as it is independent of even- or odd-numbered lattices.

\begin{figure}[!htb]
        \centering
        \includegraphics[width=\linewidth]{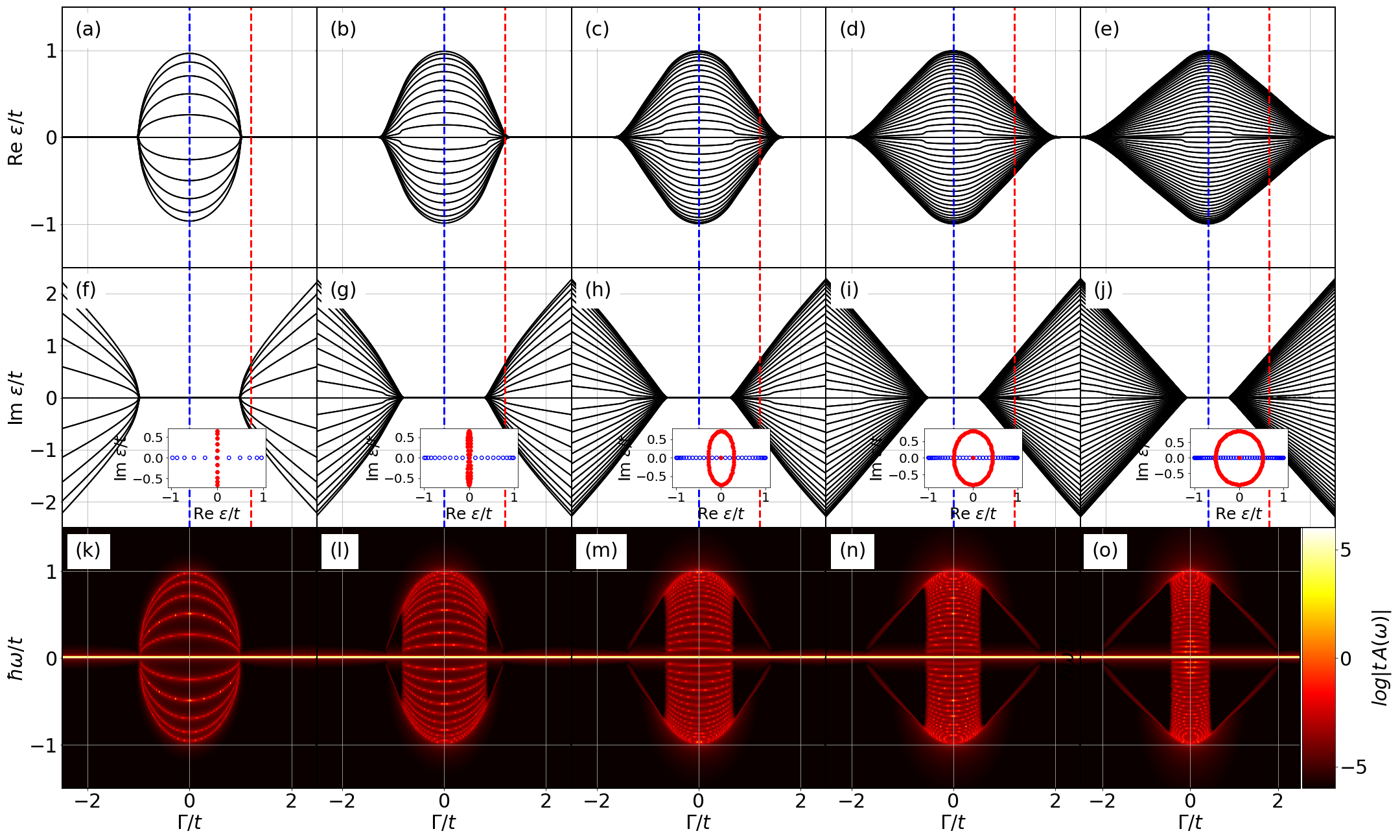}
        \caption{Spectrum and spectral weight of the OBC HNK system as a function of $\Gamma/t$ for $L=11$ (first column, (a,f,k)), $L=21$ (second column, (b,g,l)), $L=31$  (third column, (c,h,m)), $L=41$  (fourth column, (d, i,n)), and $L=51$  (fifth column, (e,j,o)). The first (second) row shows the real (imaginary) part of the spectrum, while the third row shows the log of the spectral weight $|A|$. 
      Insets in (f-j) show the energy levels in the complex plane for the system in the Hermitian limit ($\Gamma=0$, blue dashed lines and open circles) and in the $\mathcal{PH}$-broken phase ($\Gamma=1.2 t$, red dashed lines and filled circles). Other parameters: $\Delta=10^{-10} t$ and $\mu=0$.}
        \label{fig_spectrum_Delta_L_odd}
    \end{figure}

In Fig.~\ref{fig_E_Delta_odd} we additionally show the minigap $\delta$, local density of states $\rho$, and superconducting pair correlations $F$ for a lattice with $L=31$ sites and $\Delta=10^{-10}t$. This figure should be directly compared with Figs.~3 and 4 of the main text that displays the same quantities and for the same parameters but for a lattice with $L=30$ sites. The behavior of $\delta$ as a function of $\Delta$ and $\Gamma$, Fig.~\ref{fig_E_Delta_odd}(a), is the same as the one shown in Fig.~3(c), as is the behavior of $\rho$, Fig.~\ref{fig_E_Delta_odd}(c) and $F$, Fig.~\ref{fig_E_Delta_odd}(e) in the $\mathcal{PH}$-broken phase. This illustrates that exceptionally enhanced topological superconductivity is present in all systems, independent of them having an even or odd number of sites.
However, in the Hermitian limit, Figs.~\ref{fig_E_Delta_odd}(b,d), the same quantities now show additional peaks at $\omega=0$ due to the spurious zero modes. We note, however, that these modes are not localized at the edges but are fully delocalized across the chain, and thus per definition \textit{not} spatially isolated MZMs. Interestingly, they are modes with a wavelength of two lattice parameters, which causes the dotted patterns seen in Figs.~\ref{fig_E_Delta_odd}(b,d).
     \begin{figure}[!htb]
        \centering
        \includegraphics[width=\linewidth]{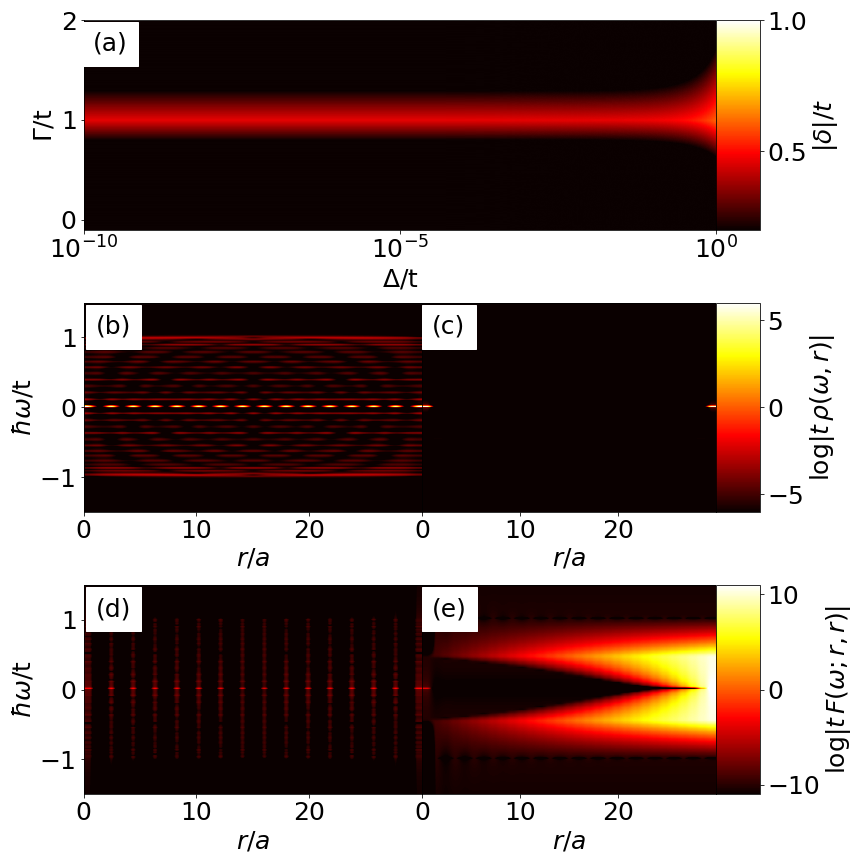}
         \caption{Properties of the MZMs and superconductivity for an odd-numbered chain. (a) Minigap $\delta$ as a function of $\Delta$ and $\Gamma$. Local density of states $\rho(\omega; r)$ as a function of $\omega$ and position $r$ along the chain ($a$ is the lattice parameter) for (b) $\Gamma=0$ and (c) $\Gamma=t$. Local superconducting pair correlations $F_{eh}(\omega; r, r)$ for (d) $\Gamma=0$ and (e) $\Gamma=t$. Other parameters: $L=31$, $\mu=0$ and $\Delta=10^{-10} t$, unless otherwise specified.}
        \label{fig_E_Delta_odd}
    \end{figure}
The analysis present in this SM shows how the system size dramatically changes the superconducting gap and spectral properties of the system due to the large degeneracy at the normal-state EP, which further supports the analysis present in the main text.

\clearpage

\section{Approximate analytical expression for spectrum at $\Gamma=t$}

In this section of the SM, we demonstrate how we obtain the analytical results in Eq.~(2) of the main text and discuss this result in more detail.  Since the HNK Hamiltonian is not a Toeplitz matrix, a generic analytic expression for the eigenvalues and eigenstates is difficult to obtain in general. However, it turns out that we can still perform an approximate analytical analysis at $\Gamma=t$.

For $\Delta=0$, $\mathbb{H}_{\rm BdG}$ is a direct sum of the electron and hole Hamiltonians $\mathbb{H}_{\rm BdG}=\left(\mathbb{H}_{e}\oplus\mathbb{H}_{h}\right)/2$. For $\Gamma=t$, both Hamiltonians assume a Jordan form
\begin{equation}
    \mathbb{H}_{e}/t=\left(\begin{matrix}0&-2&0&0&0&\cdots\\0&0&-2&0&0&\cdots\\0&0&0&-2&0&\cdots\\\vdots&\vdots&\vdots&\ddots&\ddots&\vdots\\0&0&0&0&\cdots&-2 \\0&0&0&0&\cdots&0 \\\end{matrix}\right), \qquad \mathbb{H}_{h}/t=\left(\begin{matrix}0&0&0&0&0&\cdots\\2&0&0&0&0&\cdots\\0&2&0&0&0&\cdots\\\vdots&\ddots&\ddots&\vdots&\vdots&\cdots\\0&0&\cdots&2&0&0 \\0&0&0&\cdots&2&0 \\\end{matrix}\right),
\end{equation}
typical of EPs.
As a matter of fact, the secular equation
\begin{equation}
    \det\left[\epsilon \mathbf{1}_{2L}-\mathbb{H}_{\rm BdG}\left(\Gamma=t, \Delta=0\right)\right]=\left(\epsilon\right)^{2L}=0,
\end{equation}
shows that all the $2L$ energies are equal to zero, explicitly demonstrating the very high degeneracy of the EP. The degeneracy is proportional to the number of sites $L$ and has been named an 'infernal point' in Ref.~\cite{fu2022degeneracy}.

The inclusion of even a vanishingly small value of a superconducting order parameter $\Delta$ combines the electron and hole components of the Hamiltonian. Explicitly, $\mathbb{H}_{\rm BdG}$ that was block diagonal above, now has the block form
\begin{equation}
    \mathbb{H}_{\rm BdG}=\frac{1}{2}\left(\begin{matrix}\mathbb{H}_{e}&\mathbf{\Delta}\\\mathbf{\Delta}^\dagger&\mathbb{H}_{h}\end{matrix}\right),
    \label{eq_H_BdG}
\end{equation}
where $\mathbf{\Delta}$ is the matrix of $p$-wave pairing in real space
\begin{equation}
    \mathbf{\Delta}=\left(\begin{matrix}0&\Delta&0&0&0&\cdots\\-\Delta&0&\Delta&0&0&\cdots\\0&-\Delta&0&\Delta&0&\cdots\\\vdots&\vdots&\ddots&\ddots&\ddots&\vdots\\0&0&\cdots&-\Delta&0&\Delta \\0&0&0&\cdots&-\Delta&0 \\\end{matrix}\right).
\end{equation}
Besides connecting the electron and hole components, $\mathbf{\Delta}$ makes such that $\mathbb{H}_{\rm BdG}$ is no longer a Jordan block. Although the exact expression of the determinant cannot, to the best of our knowledge, be obtained analytically, the expression of the secular equation can be expressed in terms of a series for small $\Delta$, such that we can obtain a scaling relationship for the energies. We describe this procedure in detail later, but here we first comment on the result and compare it with our numerical data. 

The secular equation can be approximated by 
\begin{equation}
    \left(\epsilon/t\right)^{2L}+(-1)^{L}\left(\epsilon/t\right)^2\left(\Delta/t\right)^2\approx 0,
    \label{eq_ep_an}
\end{equation}
which has as solutions either the MZMs with $\epsilon=0$ or the bulk states
\begin{equation}
    \epsilon_m/t\approx \exp\left(i \pi\frac{L}{2L-2} m\right)\left(\Delta/t\right)^{L-1}, 
\end{equation}
where $m=0, 1, \dots, 2L-3$ labels a different $2L-2$-th root of $(-1)^L$. These energies then describe points on a circle of radius $r=\sqrt[L-1]{\Delta/t}$ in the complex energy plane. This is the approximate analytical result reported in the main text in Eq.~(2). Although this result is similar to the response of an EP to perturbations \cite{kato2013perturbation, wiersig2022response}, it here depends on $L-1$ instead of $L$ since the modes related to the MZMs remain at zero energy. Due to the spectrum changing as $\sqrt[L-1]{\Delta/t}$, for $L\rightarrow \infty$, any $\Delta$ creates a sizable value of a gap, which is the essence of exceptionally enhanced superconductivity. 

In Fig.~\ref{fig_scaling} we compare the exact numerical results with these approximate analytical expressions for the bulk energies and also the extracted minigap to complement the results reported in the main text in Fig.~3(a,b). In Fig.~\ref{fig_scaling}(a), we compare Eq.~\eqref{eq_ep_an} with the numerical data for $\Delta=10^{-10} t$, for many different system sizes $L$. We notice that for all system sizes, there are $\epsilon=0$ solutions corresponding to the MZMs, while the bulk modes form circles in the complex plane with a radius very well described by the analytical expression in Eq.~\eqref{eq_ep_an}. There is a very minor eccentricity of these circles, but it is reduced for the larger system size, indicating that the subleading corrections to Eq.~\eqref{eq_ep_an} are only having any effect at small system sizes. The radius of the circle directly gives the minigap $\delta$ as it is the energy difference between the MZM (at zero energy) and the lowest-lying bulk modes. In Fig.~\ref{fig_scaling}(b) we plot the minigap $\delta$ as a function of $\Delta$ on a log-log plot for the same systems sizes. The dots are the values obtained by the numerical diagonalization, while dashed lines are the function $\sqrt[L-1]{\Delta/t}$. We find that the analytical expression reproduces extremely well the power-law behavior of the numerical data for not only small $\Delta$, but also larger $\Delta$ for large system sizes. Again, this means subleading corrections are not relevant for larger system sizes. 

  \begin{figure}[!htb]
        \centering
        \includegraphics[width=\linewidth]{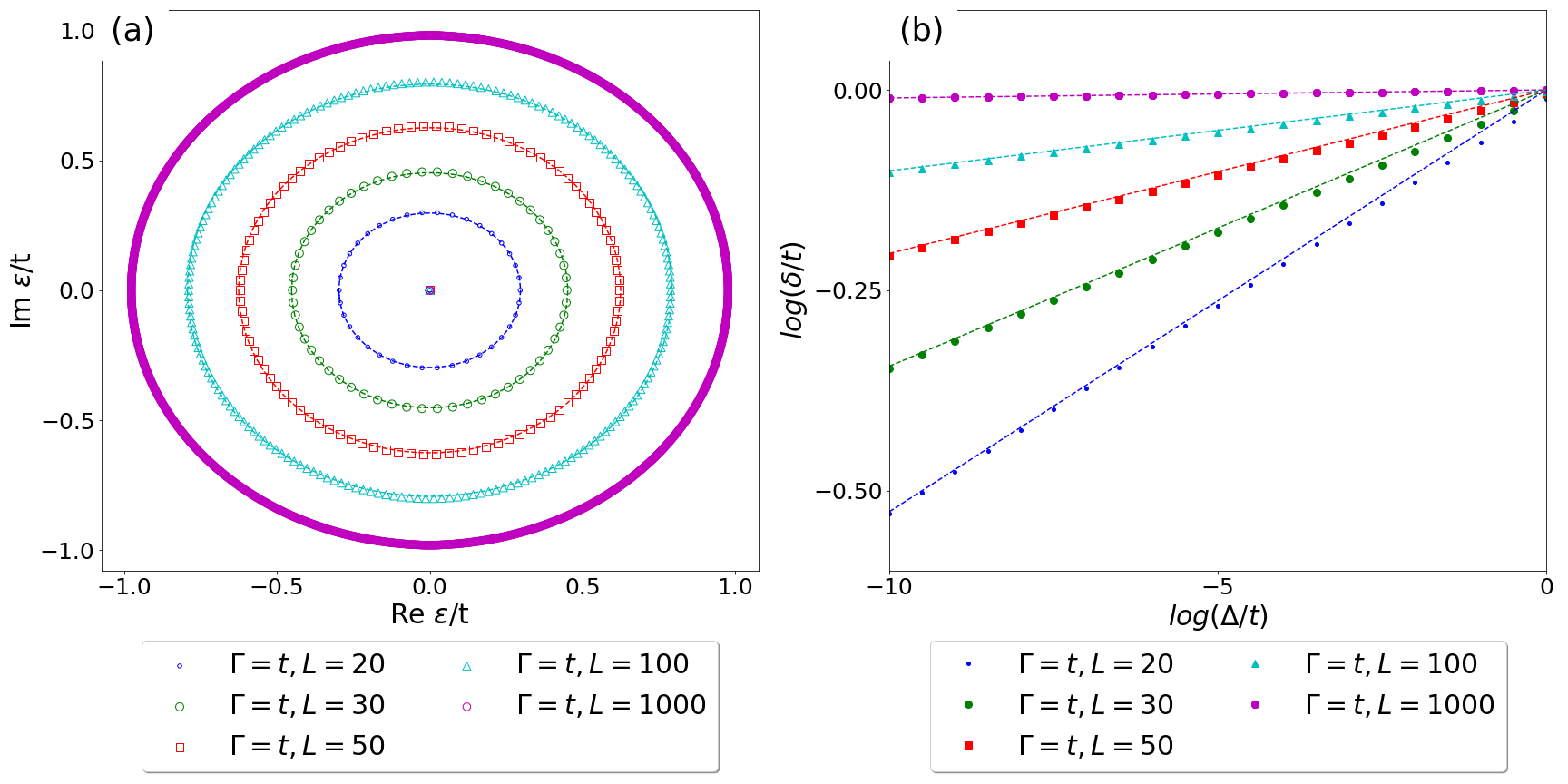}
        \caption{Comparison between the approximate analytical expressions for (a) bulk energies $\epsilon$ and (b) minigap $\delta$ and the exact numerical data for many different system sizes $L$. (a) shows $\epsilon$ in the complex plane using $\Delta = 10^{-10}t$. (b) shows the scaling of $\delta$ with $\Delta$.  Symbols indicate numerical data, dashed lines represent the analytical expressions. Other parameters: $\Gamma=t$, $\mu=0$.}
        \label{fig_scaling}
    \end{figure}

\subsection{Form of the determinant for different system sizes}

The expression of the secular determinant is complicated and, as far as we know, does not have a simple solution. Still, we can gain insight by keeping only terms of order $\Delta^2$. The resulting expression is different for even and odd numbers of sites, involving different powers of $\epsilon$, but, nevertheless, we can understand why the spectrum can be approximated by Eq.~\eqref{eq_ep_an}.

The determinant of a matrix can generally be obtained using the method of cofactors. Although this method does not generate an equation relating the determinants for different systems sizes, in contrast to what happens to Toeplitz matrices \cite{li2022exact}, we can still use this method to extract the determinant of the system for all small system sizes and from there clearly motivate the expression for any $L$. We will first perform a change of basis in $\mathbb{H}_{BdG}$ equivalent to changing the Nambu vector to $\mathbb{C}=\big(\begin{matrix}c_0&c_0^\dagger&\cdots&c_{L-1}& c_{L-1}^\dagger \end{matrix}\big)^T$, such that every two rows/columns describe the electron and holes states in each site. This change of basis is made to make the calculation of the determinant more convenient since the matrix becomes more sparse. For convenience, we also introduce the notation
\begin{equation}
    \mathcal{D}_{L}\equiv \det\left[\epsilon \mathbf{1}_{2L}-\mathbb{H}_{\rm BdG}\left(\Gamma=t, \Delta\right)\right].
\end{equation}
Below we explicitly extract $\mathcal{D}_{L}$ for $L=2,3$ followed by the expressions for large even and odd $L$.

\subsubsection{2 sites}

 We start with a system of just 2 sites. In this case, the determinant is a $4\times 4$ matrix that can be split into $3\times3$ determinants that are easily computable: 
\begin{equation}
    \mathcal{D}_2=\left|\begin{matrix}\epsilon&0&0&\Delta/2\\ 0&\epsilon&-\Delta/2&-t\\ t&-\Delta/2&\epsilon&0\\ \Delta/2&0&0&\epsilon \end{matrix}\right|=\epsilon\underbrace{\left|\begin{matrix}\epsilon&-\Delta/2&-t\\-\Delta/2&\epsilon&0\\0&0&\epsilon \end{matrix}\right|}_{\epsilon^3-\frac{\Delta^2}{4}\epsilon}-\frac{\Delta}{2}\underbrace{\left|\begin{matrix}0&\epsilon&-\Delta/2\\ t&-\Delta/2&\epsilon\\ \Delta/2&0&0\end{matrix}\right|}_{\epsilon^2 \frac{\Delta}{2}-\left(\Delta/2\right)^3}=\left(\epsilon^2-\frac{\Delta^2}{4}\right)^2.    
\end{equation} 
Thus the energies are given by $\epsilon^2=\left(\Delta/2\right)^2,$ such that $\epsilon=\pm \Delta/2$. Notice that for this small system, the modifications of the energy levels are only linear in $\Delta$.

\subsubsection{3 sites}

For three sites, the determinant is needed of a $6\times 6$ matrix, which makes its calculation cumbersome. Using the method of cofactors again, it can be split into two $5\times 5$ determinants
\begin{eqnarray}
    \mathcal{D}_3&=&\left|\begin{matrix}\epsilon&0&0&\Delta/2&0&0\\ 0&\epsilon&-\Delta/2&-t&0&0\\ t&-\Delta/2&\epsilon&0&0&\Delta/2\\ \Delta/2&0&0&\epsilon&-\Delta/2&-t\\ 0&0&t&-\Delta/2&\epsilon&0\\ 0&0&\Delta/2&0&0&\epsilon\end{matrix}\right|\nonumber\
    \\&=&\epsilon\left|\begin{matrix}\epsilon&-\Delta/2&-t&0&0\\ -\Delta/2&\epsilon&0&0&\Delta/2\\ 0&0&\epsilon&-\Delta/2&-t\\ 0&t&-\Delta/2&\epsilon&0\\ 0&\Delta/2&0&0&\epsilon\end{matrix}\right|-\frac{\Delta}{2}\left|\begin{matrix}0&\epsilon&-\Delta/2&0&0\\ t&-\Delta/2&\epsilon&0&\Delta/2\\ \Delta/2&0&0&-\Delta/2&-t\\ 0&0&t&\epsilon&0\\ 0&0&\Delta/2&0&\epsilon\end{matrix}\right|.\label{eq_3_1}
\end{eqnarray}
Again, using cofactors to split these into determinants of $4\times4$ matrices, we find the first term given by
\begin{eqnarray}
    \epsilon\left|\begin{matrix}\epsilon&-\Delta/2&-t&0&0\\ -\Delta/2&\epsilon&0&0&\Delta/2\\ 0&0&\epsilon&-\Delta/2&-t\\ 0&t&-\Delta/2&\epsilon&0\\ 0&\Delta/2&0&0&\epsilon\end{matrix}\right|&=&\epsilon^2 \underbrace{\left|\begin{matrix}\epsilon&0&0&\Delta/2\\ 0&\epsilon&-\Delta/2&-t\\ t&-\Delta/2&\epsilon&0\\ \Delta/2&0&0&\epsilon\end{matrix}\right|}_{\mathcal{D}_2}+\epsilon \frac{\Delta}{2}\underbrace{\left|\begin{matrix}-\Delta/2&-t&0&0\\ 0&\epsilon&-\Delta/2&-t\\ t&-\Delta/2&\epsilon&0\\ \Delta/2&0&0&\epsilon\end{matrix}\right|}_{-\frac{\Delta}{2}\epsilon\left(\epsilon^2-\frac{\Delta^2}{4}\right)+t^2\Delta \epsilon}\nonumber\\
    &=& \epsilon^2\left(\epsilon^2-\frac{\Delta^2}{4}\right)^2-\epsilon^2\frac{\Delta^2}{4}\left(\epsilon^2-\frac{\Delta^2}{4}\right)+\epsilon^2 t^2\frac{\Delta^2}{2},
\end{eqnarray}
while the second is given by
\begin{eqnarray}
    -\frac{\Delta}{2}\left|\begin{matrix}0&\epsilon&-\Delta/2&0&0\\ t&-\Delta/2&\epsilon&0&\Delta/2\\ \Delta/2&0&0&-\Delta/2&-t\\ 0&0&t&\epsilon&0\\ 0&0&\Delta/2&0&\epsilon\end{matrix}\right|=t\frac{\Delta}{2}\underbrace{\left|\begin{matrix}\epsilon&-\Delta/2&0&0\\ 0&0&-\Delta/2&-t\\ 0&t&\epsilon&0\\ 0&\Delta/2&0&\epsilon\end{matrix}\right|}_{\epsilon^2t\Delta} -\frac{\Delta^2}{4}\underbrace{\left|\begin{matrix}\epsilon&-\Delta/2&0&0\\ -\Delta/2&\epsilon&0&\Delta/2\\ 0&t&\epsilon&0\\ 0&\Delta/2&0&\epsilon\end{matrix}\right|}_{\epsilon^4-\epsilon^2\frac{\Delta^2}{2}}. 
\end{eqnarray}
Notice that $\mathcal{D}_3$ is not simply connected to $\mathcal{D}_2$, which indeed appears in the first term of Eq.\eqref{eq_3_1}. This is a general feature of $\mathcal{D}_L$, which illustrates the increasing complexity. 
Nevertheless, for three sites, we can collect all terms and re-express the determinant as 
\begin{equation}
    \mathcal{D}_3=\epsilon^2\left[\left(\epsilon^2-\frac{\Delta^2}{2}\right)^2+t^2\Delta^2\right],
\end{equation}
from which we obtain the energy levels $\epsilon=0$ and $\epsilon=\pm\sqrt{\Delta^2/2\pm i t\Delta}$.

Notice that the second solution changes as $\sqrt{\Delta}$ for small $\Delta$. This solution can be obtained by approximating
\begin{equation}
    \mathcal{D}_3=0=\epsilon^2\left[\left(\epsilon^2-\frac{\Delta^2}{2}\right)^2+t^2\Delta^2\right]\approx \epsilon^2\left[\epsilon^4+t^2\Delta^2\right]\rightarrow \epsilon\approx \exp\left(i \pi\frac{m}{4}\right)\sqrt{t\Delta},\quad m=0, 1, 2, 3.
\end{equation}
The same rationale can be used to obtain approximate solutions for any $L$, which show the scaling of the bulk modes with the system size. For $L>3$, there is no simple analytical solution, but this type of expansion to the lowest order in $\Delta$ motivates the scaling observed in the numerical data. 

\subsubsection{Thermodynamic limit, even $L$}
Moving beyond the analytically exact expressions for the smallest system sizes, we find that for an even $L$, the determinant $\mathcal{D}_L$ takes to the lowest order in $\Delta$ the form (checked explicitly in Mathematica up to 100 sites)
\begin{equation}
    \mathcal{D}_L=\epsilon^{2L}+\left(-\epsilon^2 t^{2L-4}+2 t^{2L-6} \epsilon^4-3 t^{2L-8} \epsilon^6+\cdots-\frac{(L-1)}{2} \epsilon^{2L-2}\right)\Delta^2+\mathcal{O}\left(\Delta^4\right).
\end{equation}
The term multiplying $\Delta^2 $ in parenthesis can be re-expressed as 
\begin{equation}
    t^{2L-2} \left[-(\epsilon/t)^2+2 (\epsilon/t)^4-3 (\epsilon/t)^6+\cdots-\frac{(L-1)}{2} (\epsilon/t)^{2L-2}\right]=t^{2L-2}\left[\sum\limits_{n=1}^{L-1}(-1)^n n \left(\epsilon/t\right)^{2n}+\frac{(L-1)}{2} (\epsilon/t)^{2L-2}\right].
\end{equation}
In the thermodynamic limit ($L\rightarrow \infty$), we can use the expression for the series
\begin{equation}
\sum\limits_{n=1}^{\infty}(-1)^n n\, x^{2n}=-\left(\frac{x}{1+x^2}\right)^2,
\end{equation}
to approximate the secular equation by
\begin{equation}
    \left(\epsilon/t\right)^{2L}-\left[\frac{\epsilon/t}{1+\left(\epsilon/t\right)^2}\right]^2\Delta^2+\frac{(L-1)}{2} (\epsilon/t)^{2L-4}=0.
    \label{eq_epsilon_an_even}
\end{equation}
This expression is valid in the thermodynamic limit $L\rightarrow\infty$ up to order $\mathcal{O}\left[\left(\Delta/t\right)^2\right]$.

\subsubsection{Thermodynamic limit, odd $L$}
For an odd $L$, we instead get a different expression for the determinant in lowest order in $\Delta$
(checked explicitly in Mathematica up to 99 sites)
\begin{equation}
    \mathcal{D}_L=\epsilon^{2L}+(t^2-\epsilon^2)\epsilon^2\left(t^{2L-6}-t^{2L-8}\epsilon^2+ 2t^{2L-10}\epsilon^4-2 t^{2L-12} \epsilon^6+\cdots+\frac{(L-1)}{2} \epsilon^{2L-6}\right)\Delta^2+\mathcal{O}\left(\Delta^4\right).
\end{equation}
Again, we can re-express
\begin{equation}
    \left(t^{2L-6}-t^{2L-8}\epsilon^2+ 2t^{2L-10}\epsilon^4-2 t^{2L-12} \epsilon^6+\cdots\right)=t^{2L-6}\sum\limits_{n=1}^{\frac{L-3}{2}} n \left[\left(\epsilon/t\right)^{4(n-1)}-\left(\epsilon/t\right)^{4n-2}\right].
\end{equation}
Notice that this expression does not include the term proportional to $\epsilon^{2L-6}$.
In the thermodynamic limit, we can use the expression of the series
\begin{equation}
    \sum\limits_{n=1}^{\infty} n \left(x^{4(n-1)}-x^{4n-2}\right)=\frac{1}{\left(1-x^2\right)(1+x^2)^2},
\end{equation}
to finally arrive at the equation
\begin{equation}
    \left(\epsilon/t\right)^{2L}+\left[\frac{\epsilon/t}{1+\left(\epsilon/t\right)^2}\right]^2\Delta^2+\frac{(L-1)}{2} (\epsilon/t)^{2L-6}=0.
    \label{eq_epsilon_an_odd}
\end{equation}
This expression is valid in the thermodynamic limit $L\rightarrow\infty$ up to order $\mathcal{O}\left[\left(\Delta/t\right)^2\right]$.

\subsubsection{Scaling and MZMs}
From the results above, we see that to the lowest order of $\Delta$ we arrive at complicated polynomial equations, which in the thermodynamic limit can be approximated by the equations, \eqref{eq_epsilon_an_odd} and \eqref{eq_epsilon_an_even}, for even and odd $L$, respectively. We notice that these two equations have two important similarities. First, both present two solutions with $\epsilon=0$. This is exactly the MZMs. Second, the scaling of $\epsilon$ with system size can be obtained by taking the log of Eqs.~\eqref{eq_epsilon_an_even} and \eqref{eq_epsilon_an_odd}. In both cases, assuming $\epsilon$ to be of form $\exp(i\theta)\Delta^x$, we obtain
\begin{equation}
\left(2L-2\right)x\log(\Delta)+i \left(2L-2\right)\theta=2\log(\Delta)+im\pi L +\log\left(\frac{1}{1+\Delta^{2x}}+(-1)^{L}\frac{L-1}{2}\Delta^{y}\right),
\label{eq_scaling}
\end{equation}
where $y=(2L-10)x$ or $y=(2L-8)x$ for odd or even $L$, respectively, and $m$ describes a branch of the log of $(-1)^L$. We thus have approximately the solutions
\begin{equation}
    x=\frac{1}{L-1}, \theta=\pi \frac{L}{2L-2}m,
\end{equation}
with $m$ restricted to be between $0$ and $2L-3$ to give the different $2L-2$ bulk solutions of the Hamiltonian. We here remark that the values of the exponents and the phases obtained in this analysis can also be obtained by just disregarding the terms in the right-hand side of Eq.~\eqref{eq_scaling} in the secular equation, which leads to Eq.~\eqref{eq_ep_an}.

To summarize, the analytical approximation explicitly shows how the enhancement of superconductivity occurs due to the large degeneracy at the normal-state EP, such that, for a large enough system, even a vanishingly small value of $\Delta$ is sufficient to open a gap of the order of $t$, as discussed in section \textit{Exceptional gap} of the main text.

 \clearpage

\section{Finite chemical potential}
Most results in the main text are reported at a chemical potential $\mu/t = 0$.
The inclusion of a finite chemical potential $\mu/t$ eventually destroys the topological phase in the Hermitian Kitaev chain. 
In particular, for $\left|\mu/t\right|>2$, the system is a topologically trivial superconductor, regardless of the values of $\Delta$ and $\Gamma$, as shown by topological invariants in Ref.~\cite{cao2021universal}. For finite $\mu$, but still within the topological phase, there are also some changes. We illustrate this behavior in  Fig.~\ref{fig_spectrum_Delta_mu} by plotting the spectrum and spectral weight for $\Delta=10^{-10}t$ with finite $\mu$, increasing from the left to right. We see that for $\mu=0.5t$, Figs.~\ref{fig_spectrum_Delta_mu}(a,f,k), the real and imaginary part of the spectrum changes in comparison with the $\mu=0$ case in Fig.~2(b, d) and Fig.~4(d) of the main text, but the system still present a point gap with isolated MZMs, which are also clearly visible in the spectral weight as a ZBP. This illustrates explicitly that all our results in the main text are stable in a substantial parameter regime centred around $\mu=0$. Even for $\mu=t$, Figs.~\ref{fig_spectrum_Delta_mu}(b,g,l), we still have MZMs for a small $\Gamma$ region around the normal-state EP, but the structure of the energies in the complex plane now starts to deviate considerably in the $\mathcal{PH}$-broken phase. There is now also an apparent breaking of particle-hole symmetry, seen in the different structures of modes with negative and positive real parts of the energy.
It is not until $\mu=1.5t$, Figs.~\ref{fig_spectrum_Delta_mu}(c,h,m), that there no longer exist  values of $\Gamma$ where a ZBP occurs in the spectral weight. Thus, we find evidence of exceptionally enhanced superconductivity for values of the chemical potential all the way up to $|\mu/t| \approx t$, although the regime of $\Gamma$-values for which this enhancement occurs decreases for increasing $\mu$. 
At $\mu=2t$, the point of the Hermitian topological phase transition, Figs.~\ref{fig_spectrum_Delta_mu}(d,i,n) shows how the system is now gapless only in the Hermitian limit ($\Gamma=0$). The spectral weight also only contains signatures of the spectrum of the system in the $\mathcal{PH}$-preserved phase. Finally, at $\mu=2.5t$, Figs.~\ref{fig_spectrum_Delta_mu}(e,j,o), there is a full energy gap with no midgap states. In SV4, we continuously vary $\mu$ for $\Delta=10^{-10} t$ to more clearly show the effect of $\mu$ in the spectrum and spectral weight.

    \begin{figure}[!htb]
        \centering
        \includegraphics[width=\linewidth]{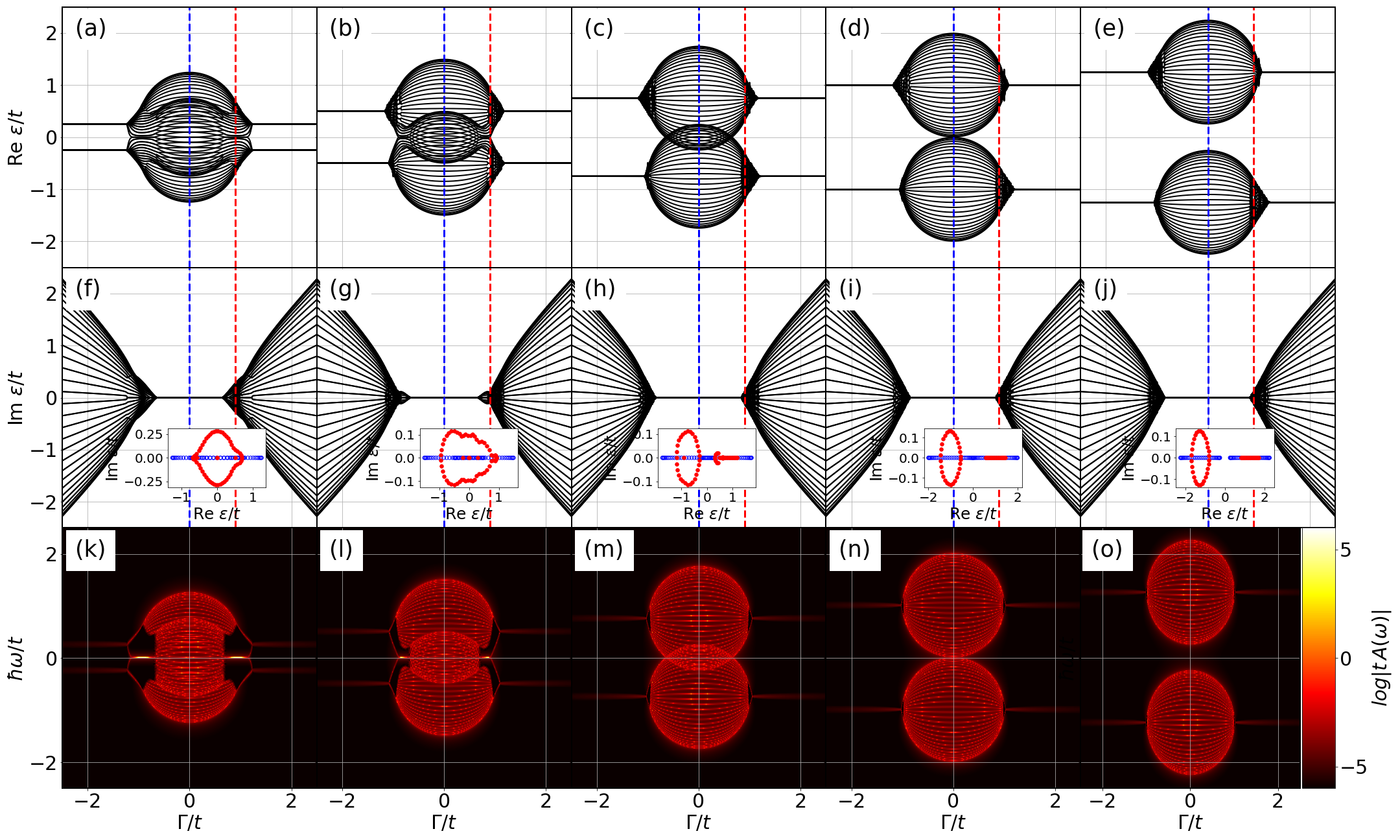}
        \caption{Spectrum and spectral weight of the OBC HNK system as a function of $\Gamma/t$ for $\mu=0.5 t$ (first column, (a,f,k)), $\mu=t$ (second column, (b,g,l)), $\mu=1.5 t$  (third column, (c,h,m)), $\mu=2.0 t$  (fourth column, (d,i,n)), and $\mu=2.5 t$ (fifth column, (e,j,o)). First (second) row shows the real (imaginary) part of the spectrum, while third row shows the log of the spectral weight $|A|$. Insets in (f-j) show the energy levels in the complex plane for the system in the Hermitian limit ($\Gamma=0$, blue dashed lines and open circles) and in the $\mathcal{PH}$-broken phase ($\Gamma=0.9 t$, red dashed lines and filled circles). Parameters: $L=30$ and $\Delta=10^{-10} t$.}
        \label{fig_spectrum_Delta_mu}
    \end{figure}
    
Finally, for completeness, we consider the effect of $\mu$ on the spectrum and spectral weight for a more sizable $\Delta$. In Fig.~\ref{fig_spectrum_Delta_mu_str}, we show the spectrum and spectral weight for increasing $\mu$ for the large value of $\Delta=t$, where also the Hermitian Kitaev chain has clearly defined MZMs. We see then that for all $\left|\mu/t\right|\neq 2$, the system presents a line gap, see insets of Figs.~\ref{fig_spectrum_Delta_mu_str}(f-j) and hosts MZMs, while for $\left|\mu/t\right|>2$ there are no midgap states in the spectral function. This is precisely the phase diagram of the HNK system, also obtained in Ref.~\cite{cao2021universal}. In SV5, we continuously change $\mu$ for $\Delta=t$ to show its effect more clearly.

    \begin{figure}[!htb]
        \centering
        \includegraphics[width=\linewidth]{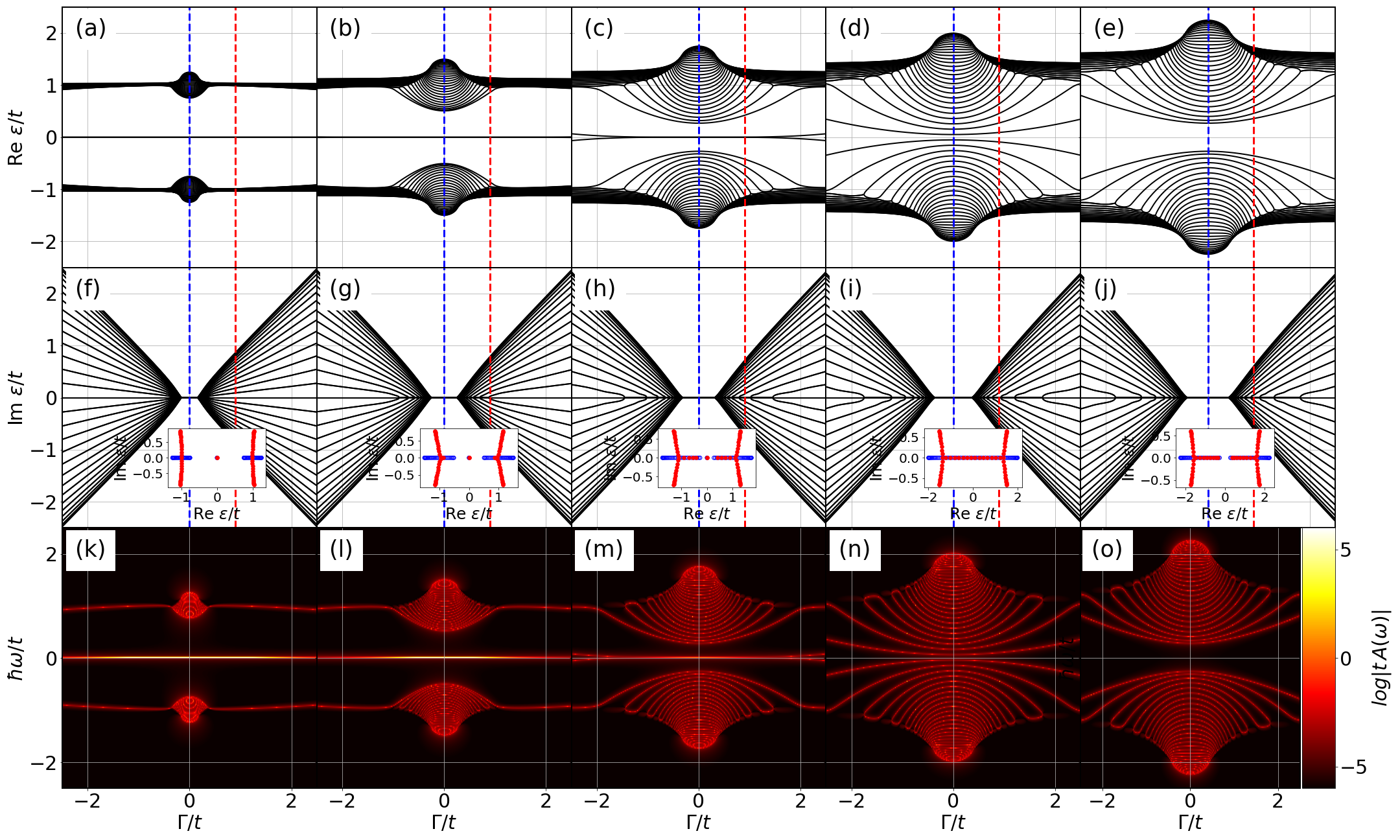}
        \caption{Spectrum and spectral weight of the OBC HNK system as a function of $\Gamma/t$ for $\mu=0$ (first column, (a,f,k)), $\mu=t$ (second column, (b,g,l)), $\mu=1.5 t$  (third column, (c,h,m)), $\mu=2.0 t$  (fourth column, (d,i,n)), and $\mu=2.5 t$ (fifth column, (e,j,o)). First (second) row shows the real (imaginary) part of the spectrum, while third row (k-o) shows the log of the spectral weight $|A|$. Insets in (f-j) show the energy levels in the complex plane for the system in the Hermitian limit ($\Gamma=0$, blue dashed lines and open circles) and in the $\mathcal{PH}$-broken phase ($\Gamma=1.2 t$, red dashed lines and filled circles). Parameters: $L=30$ and $\Delta= t$.}
        \label{fig_spectrum_Delta_mu_str}
    \end{figure}
    
In summary, the exceptionally enhanced superconductivity due to the normal-state EP is slowly diminishing with increasing $\mu$, but still persists all the way up to $|\mu/t| \approx 1$. This shows that our results in the main text are not fine-tuned to $\mu =0$. With increasing $\Delta$, we find MZMs also for larger $\mu/t$, but this is then probing the straightforward non-Hermitian extension of the Hermitian Kitaev chain. 

\clearpage

\section{Density of states, local density of states, and superconducting amplitudes for non-Hermitian systems}

In this section of the SM, we derive the expressions used for the spectral weight $A$, the local density of states $\rho$, and superconducting amplitudes $F$ in the main text. These properties are all encoded in the Green's function and many references already discuss their use, such as \cite{feinberg1997non, kozii2017non, sukhachov2020non, cayao2022exceptional}, with the use in non-Hermitian systems being very similar to the use in Hermitian systems. 
As a matter of fact, as discussed, for instance, in Ref.~\cite{datta1997electronic}, a consistent discussion of self-energy effects needs to always consider that the inverse of a Green's function is a non-Hermitian operator. 
For completeness, we provide below the relevant summary of already known results, targeted to the need in the main text.

The spectral weight $A(E)$ quantifies the weight of quasiparticles at energy $E$. For non-interacting systems, it is equivalent to the density of states, which quantifies the number of states at a given energy $E$. For a Hermitian system, this can be expressed as
\begin{equation}
    A(E)=\sum_{m}\delta\left(\epsilon_m-E\right).
\end{equation}
Using the (retarded) Green's operator 
\begin{equation}
    \hat{G}(E)=\frac{1}{\left(E+i \eta\right)\hat{1}-\hat{H}},
\end{equation}
we can express, in the limit that $\eta\rightarrow 0$,
\begin{equation}
    -\frac{1}{\pi}\lim\limits_{\eta\rightarrow 0}\text{Im} \text{Tr}\left[\hat{G}(E)\right]=\sum_{m}\delta\left(\epsilon_m-E\right)=A(E).
\end{equation}
For a non-Hermitian system, the spectral weight $A$ is now given by a combination of the retarded and advanced (Hermitian conjugate) Green's operators \cite{feinberg1997non, kozii2017non}
\begin{equation}
    A(E)=-\frac{1}{2\pi}\lim\limits_{\eta\rightarrow 0}\text{Im} \text{Tr}\left[\hat{G}(E)-\hat{G}^\dagger(E)\right]=-\frac{1}{2\pi}\lim\limits_{\eta\rightarrow 0}\text{Im} \text{Tr}\left[\hat{G}(E)-\hat{G}{^\dagger}(E)\right],
\end{equation}
such that we still obtain a positive $A(E)$. Assuming $E$ to be real, of the form $\hbar \omega$, and that the spectrum at most comes in complex conjugated pairs, as always the case for the HNK chain, we have 
\begin{align}
    \begin{split}
        \text{Tr}\left[\hat{G}(\hbar \omega)-\hat{G}^{\dagger}(\hbar \omega)\right]&=\sum\limits_{m} \left(\frac{1}{\hbar \omega+i \eta-\epsilon_m}-\frac{1}{\hbar \omega-i \eta-\epsilon_m^*}\right)\\
        &=2\sum\limits_{m'} \frac{\hbar\omega +\epsilon_{m'}-i \eta}{\left(\hbar \omega-\epsilon_{m'}\right)^2+\eta^2}+4\sum\limits_{m''} \frac{\hbar\omega +\text{Re }\epsilon_{m''}-i \eta}{\left(\hbar \omega-\text{Re }\epsilon_{m''}\right)^2+\left(\eta-\text{Im }\epsilon_{m''}\right)^2},    
    \end{split}
\end{align}
where $m'$ labels the energy levels with zero imaginary part and $m''$ labels the pair of levels that come in complex conjugated pairs. Notice that we here assume that $\eta\ll 1$ to arrive at the second term in the last equality.
As a consequence, we find
\begin{equation}
    A(\hbar \omega)= -\frac{1}{2\pi}\lim\limits_{\eta\rightarrow 0}\text{Im} \text{Tr}\left[\hat{G}(\hbar\omega)-\hat{G}^{\dagger}(\hbar\omega)\right]=\sum\limits_{m'}\delta(\hbar\omega-\epsilon_{m'}),
\end{equation}
which only has finite support for states with purely real energy. 

%We note that for a spectrum that is not PH, what is obtained also considering finite $\eta$, one needs to consider that in general $\hat{G}\left(\hbar\omega\right)\propto \prod_m (\hbar\omega-\epsilon_m)^{-1}$ such that this interpretation is still valid with the imaginary part of the energy measuring the weight of each level.

From the Green's operator, we can obtain the Green's function by projecting to a specific basis. Considering that we have a superconductor defined on a lattice, we can write
\begin{equation}
   \braket{v, r|\hat{G}(\hbar\omega)|v', r'}=\frac{1}{\left(\hbar \omega +i\eta\right)\delta_{v, v'}\delta_{r,r'}-\mathbb{H}_{v, r;v', r'}}\equiv \mathcal{G}(\omega)_{v, r; v', r'},
\end{equation}
where $v$ represents electron ($e$) or hole ($h$) components, and the subscripts on $\mathbb{H}$ and $\mathcal{G}$ refer to matrix components. Thus we arrive at
\begin{equation}
    A(\hbar \omega)= -\frac{1}{2\pi}\lim\limits_{\eta\rightarrow 0}\text{Im} \text{Tr}\left[\hat{G}(\hbar\omega)-\hat{G}^{\dagger}(\hbar\omega)\right]=-\frac{1}{\pi}\lim\limits_{\eta\rightarrow 0}\text{Im}\sum\limits_{r}\left[G_e(\omega)_{r; r}+G_h(\omega)_{r; r}\right]\equiv \sum\limits_{r}\rho_e(\omega;r)+\rho_h(\omega;r),
\end{equation}
where $\rho_{e/h}$ is the electron/hole local density of states. Notice that in the main text, we used $\rho=\rho_e$ due to the convention that the local density of states is related to the occupation of electrons only. We see that the local density of states shows the occupation of the electronic levels for a given frequency.
Using the above equations and noting that for non-Hermitian systems the resolution of the identity is written using the right $\ket{m}^R$ and left $\ket{m}^L$ eigenstates of $H$, we arrive at
\begin{align}
    \begin{split}
        \braket{v, r|\hat{G}(\hbar\omega)|v', r'}&=\bra{v, r}\hat{G}(\hbar\omega)\left(\sum\limits_{m}\ket{m}^{R\,\,\,\, L}\bra{m}\right)\ket{v', r'}=\sum\limits_{m}\braket{v, r|\frac{1}{\left(\hbar\omega+i \eta\right)\hat{1}-\hat{H}}|m}^{R\,\,\,\, L}\braket{m|v', r'}\\
        &=\sum\limits_{m}\frac{\psi_m^R(v, r)\psi_m^{L*}(v', r')}{\left(\hbar\omega+i \eta\right)-\epsilon_m}\; \; \; \therefore \; \; \; \rho(\omega;r)=\sum\limits_{m'}\text{Re} \left[\psi_{m'}^R(e, r)\psi_{m'}^{L*}(e, r)\right]\delta\left(\hbar \omega-\epsilon_{m'}\right),
    \end{split}
\end{align}
where we use again that we have a complex conjugated spectrum. We note that this is similar to the definition of the local density of states in a Hermitian system, but with the difference that $\psi^R\neq \psi^L$ in general.

For the superconducting pair amplitudes, we simply use the anomalous components of $\mathcal{G}$, resulting in
\begin{align}
    \begin{split}
        F_{eh}(\omega;r)\equiv \braket{e, r|\hat{G}(\hbar\omega)|h, r}&=\sum\limits_{m'}\frac{\psi_{m'}^R(e, r)\psi_{m'}^{L*}(h, r)}{\left(\hbar\omega+i \eta\right)-\epsilon_m}.
    \end{split}
\end{align}
Note that this expression does not become a $\delta$ function, but it is still highly peaked on states with real energies.

As a final remark, we notice that the MZMs present distinguished peaks in all the above quantities since they are fully real. We also note that $F$ presents an asymmetry between left and right edges due to the fact that the electron component of $\psi^R$ and the hole component of $\psi^L$ are localized on the same edge due to the non-Hermitian skin effect.

\end{onecolumngrid}

\end{document}